 \documentclass[serif,twocolumn]{wiley-article}
 \usepackage[super,comma,sort&compress]{natbib}
 \makeatletter
 \renewcommand{\@biblabel}[1]{#1.}
 \makeatother
 
\usepackage{siunitx}
\usepackage{tikz}
\usepackage{xcolor}
\usepackage{bm}
\usepackage{physics}
\usepackage{tabularx}
\usepackage{wasysym}
\usepackage{graphicx}
\usepackage{dcolumn}
\usepackage{bm}
\usepackage{tabularx}
\usepackage{flafter}
\usepackage{wasysym}
\usepackage{xcolor}
\usepackage{ulem}
\definecolor{darkgreen}{rgb}{0.0, 0.5, 0.0}

\papertype{Original Article}
\paperfield{Journal Section}

\title{Designing minimally-segregating granular mixtures for gravity-driven surface flows}

\author[1]{Yifei Duan}
\author[2]{Jack Peckham}
\author[2]{Paul B. Umbanhowar}
\author[1,2,3]{Julio M. Ottino}
\author[1,2,3]{Richard M. Lueptow}

\affil[1]{Department of Chemical and Biological Engineering, Northwestern University, Evanston, Illinois 60208, USA}
\affil[2]{Department of Mechanical Engineering, Northwestern University, Evanston, Illinois 60208, USA}
\affil[3]{Northwestern Institute on Complex Systems (NICO), Northwestern University, Evanston, Illinois 60208, USA}

\corraddress{Richard M. Lueptow, Department of Mechanical Engineering, Northwestern University, Evanston, IL 60208, USA.}
\corremail{r-lueptow@northwestern.edu}

\fundinginfo{NSF Grant no. CBET-1929265.}

\begin{document}

\begin{frontmatter}
\maketitle

\begin{abstract}
In dense flowing bidisperse particle mixtures varying in size or density alone, smaller particles sink (driven by percolation) and lighter particles rise (driven by buoyancy). But when the particle species differ from each other in both size and density, percolation and buoyancy can either enhance (large/light and small/heavy) or oppose (large/heavy and small/light) each other. In the latter case, a local equilibrium condition can exist in which the two segregation mechanisms balance and particles remain mixed: this allows the design of minimally-segregating mixtures by specifying particle size ratio, density ratio, and mixture concentration. Using experimentally validated DEM simulations, we show that mixtures specified by the methodology remain relatively well-mixed in the thin rapid surface flows characteristic of heaps and tumblers commonly used in industry. Furthermore, minimally-segregating particle mixtures prepared in a fully segregated state in a tumbler mix over time and eventually reach a state of nearly uniform species concentration.

\keywords{segregation, mixing, granular flow, heaps, tumblers, non-segregating}
\end{abstract}
\end{frontmatter}

\section{Introduction}

Dense flows of granular materials tend to spontaneously segregate by particle size,\cite{savage1988particle,makse1997spontaneous,may2010shear,golick2009mixing,bhattacharya2014chute,khola2016correlations,fan2017segregation} density,\cite{khakhar1997radial,tripathi2013density,liao2014density,liu2017transport,duan2020segregation}, shape,\cite{alizadeh2017effect,pereira2017segregation,lu2020particle,jones2021predicting} friction coefficient,\cite{kondic2003segregation,gillemot2017shear} or other physical properties, which can be problematic in many industries due to the deleterious impact of inhomogeneity on product quality.\cite{standish1985studies,muzzio2002powder,ottino2008mixing,gray2010large,gray2011multi,gray2018particle,umbanhowar2019modeling} 
Among the particle properties that drive segregation, size and density are usually the dominant factors.\cite{alonso1991optimum,jain2005regimes} 
In dense flows of size-disperse equal-density particles (S-system), large particles tend to rise as small particles fall through voids,~\citep{williams1968mixing,drahun1983mechanisms,ottino2000mixing} a segregation mechanism known as percolation.
For density-disperse equal-size particle mixtures (D-system), segregation is driven by a buoyant force mechanism in which heavy particles sink and light particles rise.\citep{ristow1994particle,khakhar1999mixing,pereira2011insights} 
When particle species differ from each other in both size and density (SD-system), the two segregation mechanisms interact, resulting in more complicated segregation behavior.\citep{metcalfe1996pattern,jenkins2002segregation,felix2004evidence,tunuguntla2014mixture,larcher2015evolution,jing2020rising} Though size and density differences can reinforce each other, i.e., in mixtures of large light particles and small heavy particles, we are interested here in the opposite situation where the two segregation mechanisms oppose each other, i.e., in mixtures of large heavy particles and small light particles, as this case has the potential to reduce species segregation compared to the corresponding pure S- or pure D-system.

Previous studies show that the tendency of spherical particles to sink or rise in a bidisperse mixture can be characterized by the ratios of large to small particle diameter, $R_\mathrm{d}=d_\mathrm{l}/d_\mathrm{s}$ (subscript $\mathrm{l}$ for large
particles and $\mathrm{s}$ for small particles), and density, $R_\mathrm{\rho}=\rho_\mathrm{l}/\rho_\mathrm{s}$, along with the mixture volume concentration $c_\mathrm{l}$ (or equivalently $c_\mathrm{s}$, as $c_\mathrm{l}$+$c_\mathrm{s}=1$, assuming the solid volume fraction $\phi$ is constant).\cite{gray2005theory,tunuguntla2017comparing,fan2014modelling,schlick2015granular,jones2018asymmetric}
Unlike size or density segregation alone, where which species rises or sinks depends only on $R_\mathrm{d}$ or $R_\mathrm{\rho}$ but  does not depend on mixture concentration, the segregation direction in an SD-system can be concentration dependent, and a mixed equilibrium state can exist where the effects of particle size and density differences balance.\cite{alonso1991optimum,gray2015particle}

Although granular systems are challenging to study analytically compared to fluid flows because governing equations analogous to the Navier-Stokes equations for fluids are not yet fully established (although progress is being made in this direction\cite{jenkins1983theory,gdr2004dense,pouliquen2006flow,jop2006constitutive,kamrin2012nonlocal,henann2013predictive}), granular materials can offer a conceptual advantage compared to fluids.  In many systems of practical interest---particularly tumbling and heap formation---the granular flow occurs only in thin regions of rapid flow,\cite{meier2007dynamical} even in large-scale industrial processes.  Thus, understanding flow and segregation in this thin shear layer provides a key building block for understanding how to scale-up processes of industrial significance.\cite{ottino2002scaling}  
In fact, we recently developed a segregation model based on the continuum advection-diffusion equation\cite{dolgunin1995segregation,dolgunin1998development,gray2018particle,umbanhowar2019modeling} that predicts the degree to which two non-cohesive particle species differing in both size and density will segregate in thin rapid  surface flows.\cite{duan2021modelling}
Specifically, the model predicts a segregation velocity that depends linearly on the local shear rate and quadratically on the species concentration.
The segregation velocity is characterized by two empirical coefficients that are functions of $R_\mathrm{d}$ and $R_\mathrm{\rho}$. 
Streamwise concentration profiles predicted by incorporating this segregation velocity model into a continuum advection-diffusion-segregation transport model match DEM simulation results well for free surface heap flows over a wide range of $R_\mathrm{d}$ and $R_\mathrm{\rho}$.
An important feature of the model is the ability to predict the large particle ``equilibrium concentration,'' $c_\mathrm{l,eq}$, at which  size-related percolation and  density-related buoyancy offset one another such that the segregation flux of each of the two species is zero.\citep{duan2021modelling,duan2021designing}  
In other words, an initially mixed bidisperse particle mixture does not segregate.

In this paper, we use model predictions for $c_\mathrm{l,eq}$ to demonstrate how minimally-segregating granular mixtures can be designed based on appropriate choices of $R_\mathrm{d}$, $R_\mathrm{\rho}$, and $c_\mathrm{l}$. Particle mixtures prepared near the predicted equilibrium conditions are tested in discrete element method (DEM) simulations for free surface flows on bounded heaps and in rotating tumblers to characterize the degree to which the particles remain mixed. We also provide a limited number of experiments confirming the rotating tumbler simulations.  Because flows of granular materials are often restricted to thin regions of rapid surface flow even in large-scale systems, this research offers an approach to intentionally design particle systems for industrial processes for which otherwise segregating particles will become or remain relatively well-mixed.

\begin{figure}[t]
\centering
\includegraphics[width=\columnwidth]{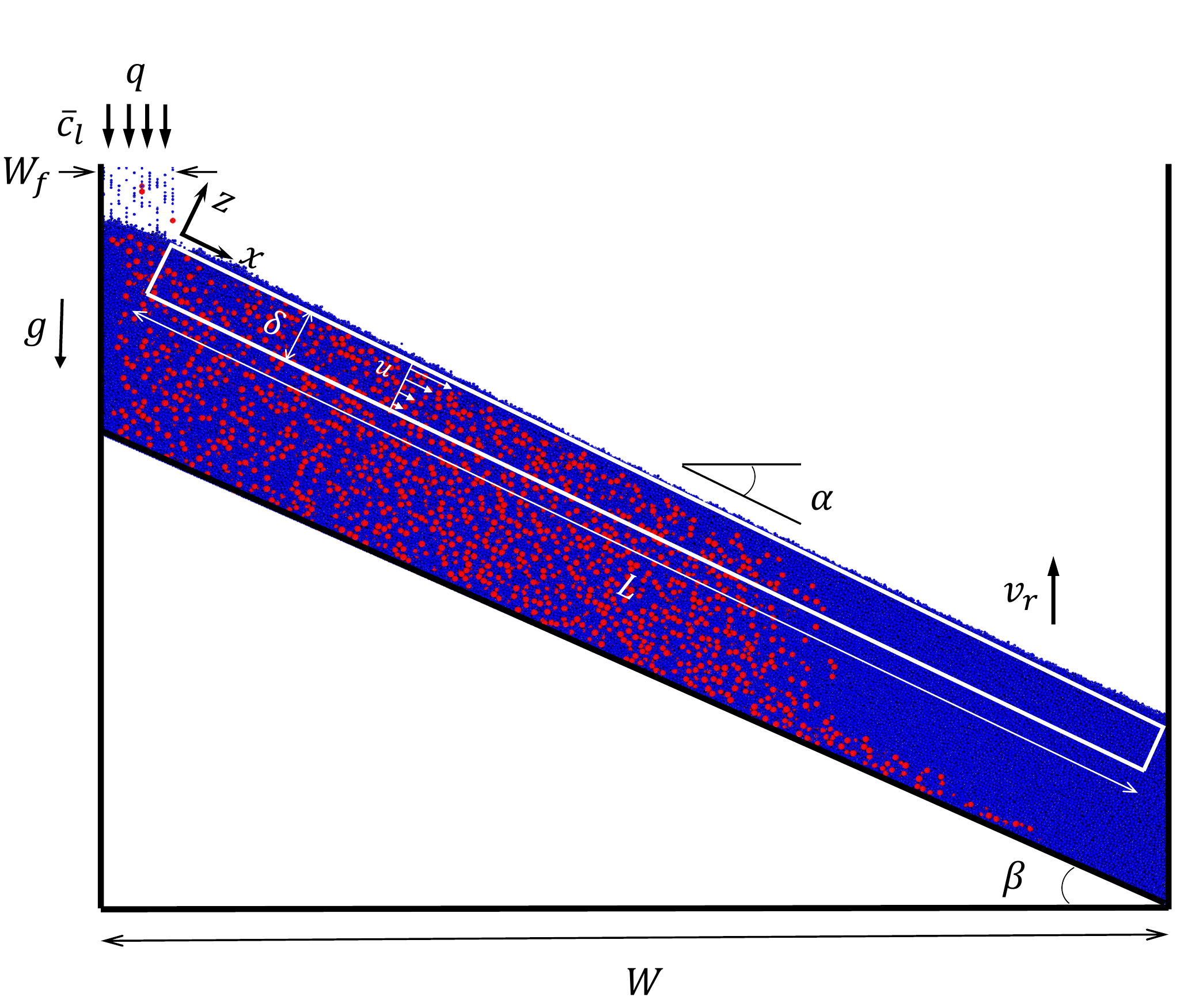}
 \caption{ Quasi-2D bounded heap setup and segregation example from DEM simulation.
 For a large-particle feed concentration of $\bar c_\mathrm{l}=0.2$, large heavy particles (\textcolor{red}{red}, $d_\mathrm{l}=3\,$mm, $\rho_\mathrm{l}=4\,$g/cm$^3$) sink, while small light particles (\textcolor{blue}{blue}, $d_\mathrm{s}=1.5\,$mm, $\rho_\mathrm{s}=1\,$g/cm$^3$) rise in the flowing layer producing a large-particle enriched upstream region and a nearly pure-small-particle downstream region in the static portion of the heap.
 $R_\mathrm{d}=2$, $R_\mathrm{\rho}=4$, $W_f=3.3\,$cm, $W=50\,$cm, $L=52\,$cm, $q=20\,$cm$^2/$s, $\delta \approx 1.5\,$cm, $v_r=0.4\,$cm/s, $\alpha=26.3^\circ$.  }
 \label{fig1}
\end{figure}

\section{Predicting Non-segregating Mixture Conditions}

\begin{figure}[t]
\centering
\includegraphics[width=\columnwidth]{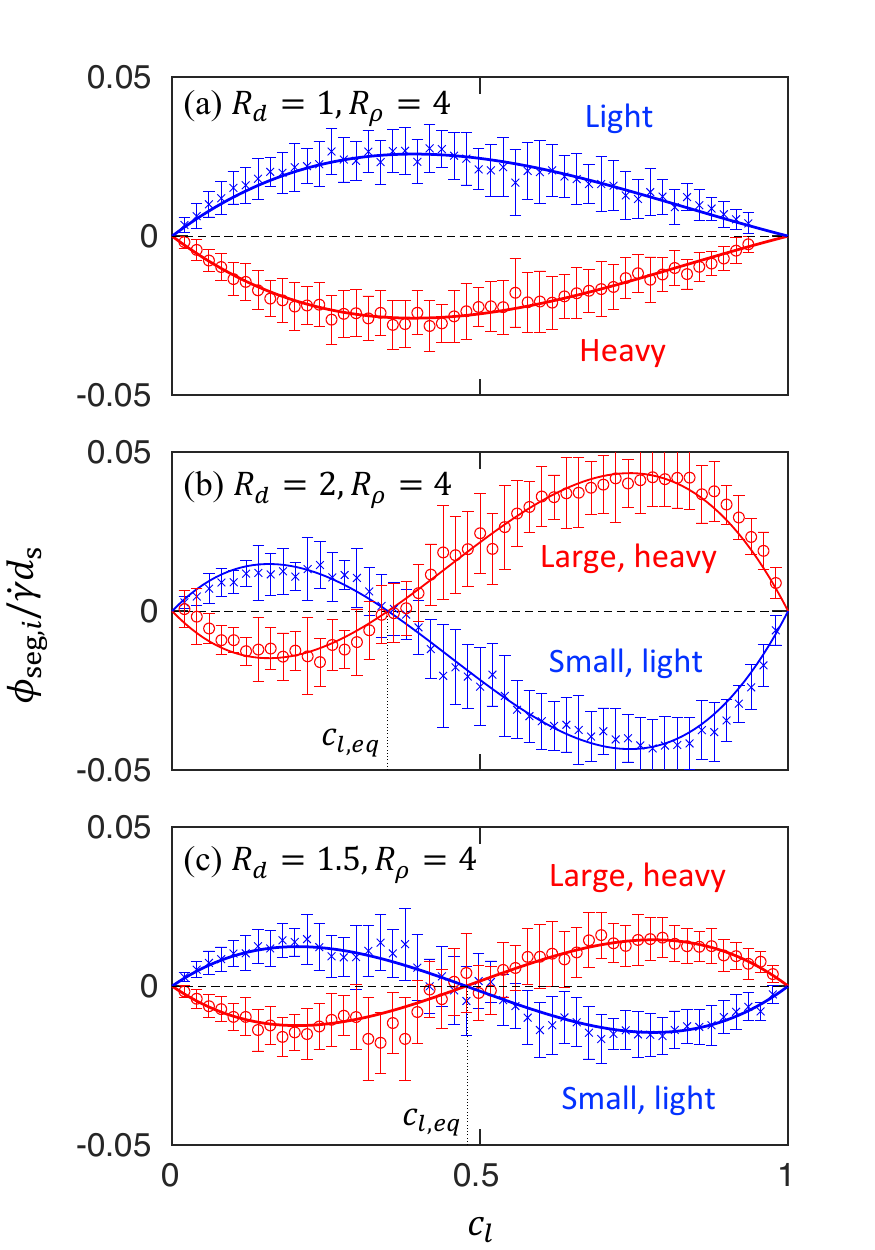}
\caption{Non-dimensionalized segregation flux data, $\phi_{\mathrm{seg},i}/\dot \gamma d_s$, for large ($\Circle$) and small  particles ($\times$)  averaged over 0.02 wide increments of $c_\mathrm{l}$ from heap flow simulations. 
Error bars represent the standard deviation for each averaging interval of $c_\mathrm{l}$.
Solid curves are fits of the segregation velocity model, Equation~(\ref{eq1}).\cite{duan2021modelling}}
\label{fig2}
\end{figure}

Consider DEM simulations of combined size and density segregation of a bidisperse mixture in a single-sided quasi-2D bounded heap flow. For the example shown in Figure~\ref{fig1}, the domain has width $W$ and thickness $T$ in the $y$-direction. Particles flow down and to the right in a thin flowing layer  of length $L$ and relatively constant thickness $\delta$ \cite{fan2013kinematics} (corresponding approximately to the white rectangle having a depth exaggerated by a factor of about two to make it more visible) and are deposited continuously at a uniform rate onto the static bed. Particles are fed onto the left side of the heap at an effective 2-D flow rate $q=Q/T$ ($Q$ is the volumetric feed rate) and with large particle feed concentration $\bar c_\mathrm{l}$, where the bar indicates the feed concentration or the global concentration rather than the local large particle concentration, $c_\mathrm{l}$; the heap rises with velocity $v_\mathrm{r}=q/WT$. 
The coordinate system is rotated by the repose angle, $\alpha$, such that the $x$-axis is in the steamwise direction, the $y$-axis is in the spanwise direction, and the $z$-axis is normal to the free surface. 
 The origin is located on the free surface at the downstream edge of the vertical feed region and rises with the free surface of the heap. 
To reduce computation time, the bottom wall is inclined at an angle $\beta= 28^\circ$, roughly matching the repose angle $\alpha$. 
Simulations are performed using our in-house DEM code,\citep{duan2021modelling,isner2020axisymmetric} which runs on CUDA-enabled GPUs and has been previously validated by heap flow experiments with mm-sized particles.\citep{xiao2016modelling,isnergranular}
For all simulations, particle-particle and particle-wall contacts use a friction coefficient of $\mu=0.4$,
a binary collision time of $t_c=0.5\,$ms, and a relatively low restitution coefficient of $e=0.2$ to minimize the downstream flux of bouncing particles.

In the case shown in Figure~\ref{fig1} for large heavy particles (red) with $\bar c_\mathrm{l}=0.2$ and small light  particles (blue) with $R_\mathrm{d}=2$ and $R_\mathrm{\rho}=4$, the conditions are such that the small light particles (blue) rise to the surface of the flowing layer.  As a result, the small light (blue) particles flow further down the slope to deposit in a nearly pure blue particle region at the downstream end of the heap, while large heavy particles (red) mixed with the small light particles (blue) deposit on the upstream portion of the heap.

The advantage of considering heap flows over other flow configurations (e.g., plane shear flows or chute flows) is that the local shear rate $\dot\gamma$ and the particle species concentration $c_i$ vary throughout the length and depth of the flowing layer but remain constant at a particular location in the flow (when analysed in a reference frame that rises with the heap surface at rise velocity, $v_\mathrm{r}$). As a result, the time-averaged segregation flux $\phi_{\mathrm{seg},i}$ ($\phi_{\mathrm{seg},i}=w_ic_i$ where $w_i$ is the species-specific velocity in the $z$-direction)  for a wide range of flow conditions ($\dot\gamma$ and $c_{i}$) can be obtained at different locations in the flowing layer from just one simulation. 
However, the full range of local concentrations is not usually realized in a single simulation, especially for weakly segregating mixtures, so typically several simulations are conducted with different large particle feed concentrations,  $\bar c_\mathrm{l}$, to provide data covering the full range of possible local concentrations $0\leq c_\mathrm{l}\leq1$.\cite{jones2018asymmetric, duan2021modelling}

\begin{figure}[t]
\centering
\includegraphics[width=\columnwidth]{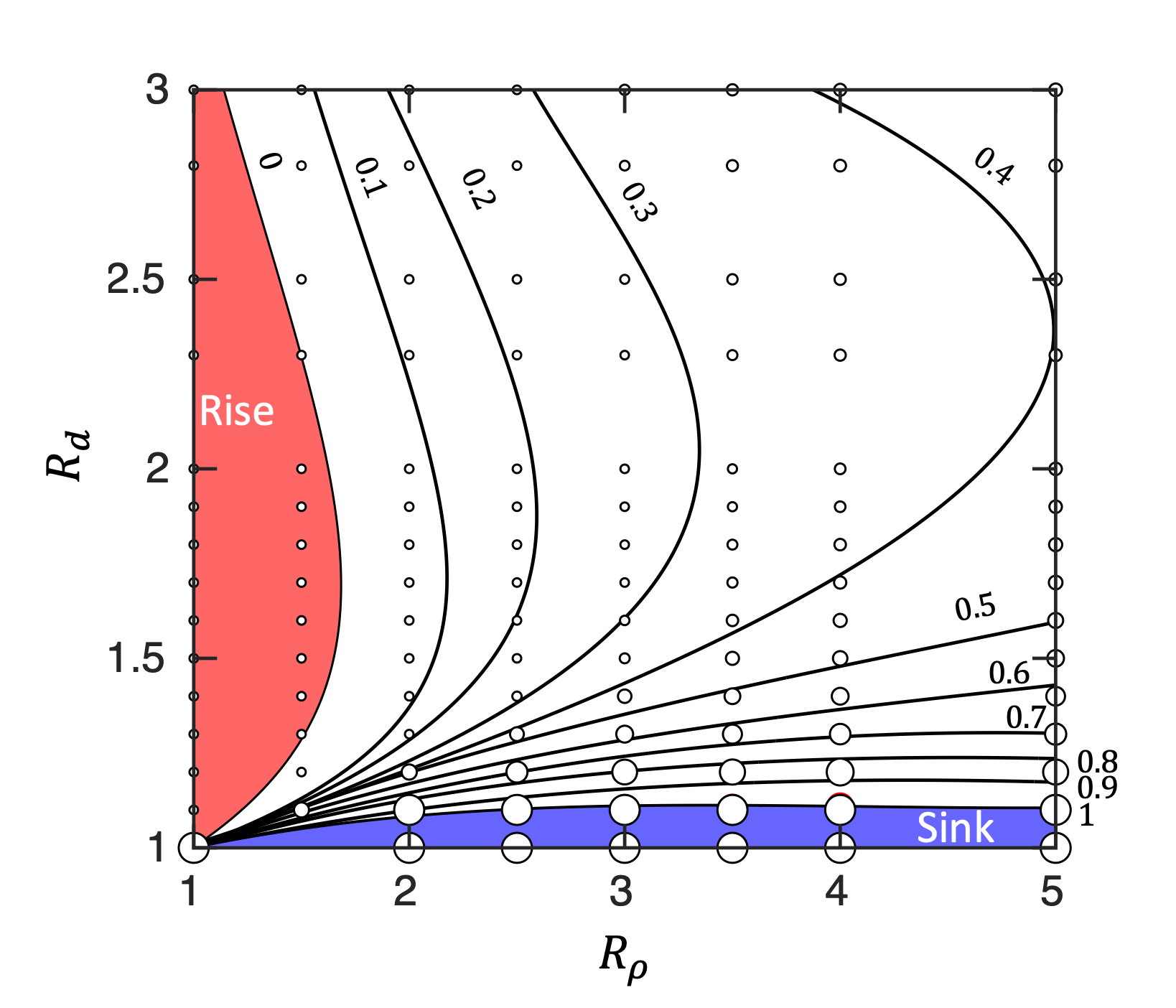}
\caption{ Local equilibrium (no segregation) concentration of large particles $c_\mathrm{l,eq}$ vs.\ particle size and density ratios. 
Circles denote combinations of $R_\mathrm{d}$ and $R_\mathrm{\rho}$ for which heap flow simulations have been performed, and circle diameter is proportional to $c_\mathrm{l,eq}$ in the range 0 to 1. Iso-concentration curves for $c_\mathrm{l,eq}$ are interpolated between data points. Particles remain mixed along the curve for $c_\mathrm{l}=c_\mathrm{l,eq}$ for the corresponding $R_\mathrm{d}$ and $R_\mathrm{\rho}$. 
Segregation is uni-directional in the filled regions for $c_\mathrm{l,eq}=0$ (large particles rise, red) and $c_\mathrm{l,eq}=1$ (large particles sink, blue).
}
\label{fig3}
\end{figure}

\begin{figure}[t]
\centering
\includegraphics[width=\columnwidth]{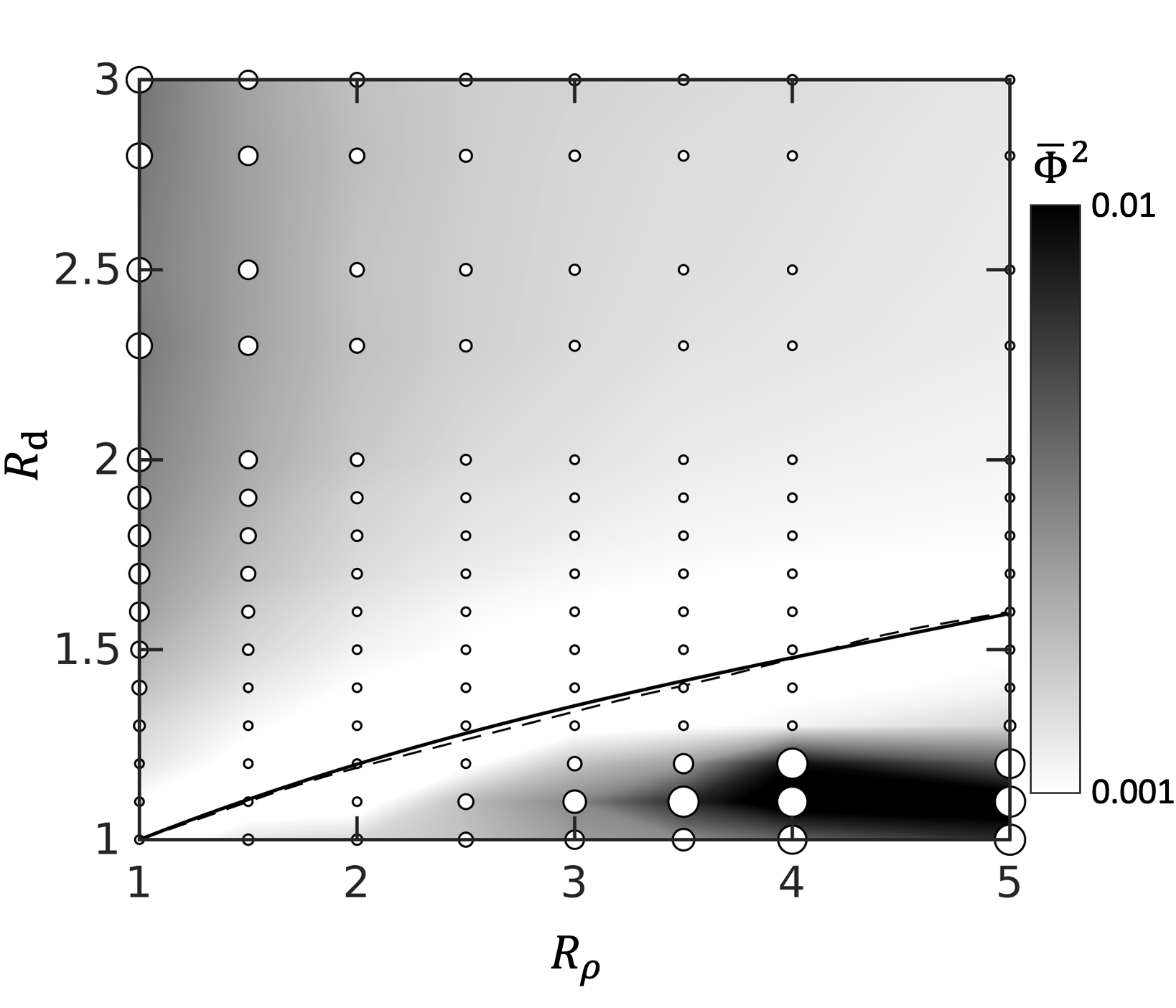}
\caption{ Concentration averaged dimensionless segregation flux squared, ${\bar\Phi}^2$, vs.\ particle size and density ratios. 
Circular symbols indicate ($R_\mathrm{d}$,$R_\mathrm{\rho}$) pairs for which simulations are performed; circle diameter is proportional to ${\bar\Phi}^2$ in the range 0.001 to 0.01.
The minimum value of ${\bar\Phi}^2$ is well-approximated by $R_\mathrm{d}=0.163(R_\mathrm{\rho}-1)+1$ (dashed curve), which nearly matches the solid curve for equilibrium concentration $c_\mathrm{l,eq}=0.5$ (reproduced from Figure~\ref{fig3}). 
}
\label{fig4}
\end{figure}

Based on simulations like that shown in Figure~\ref{fig1}, the segregation flux dependence on species concentration can be obtained for different values of $R_\mathrm{d}$ and $R_\mathrm{\rho}$ for SD-systems.
Figure~\ref{fig2} shows the mean scaled segregation flux $\phi_{\mathrm{seg},i}/\dot\gamma d_\mathrm{s}$ versus local large particle concentration $c_\mathrm{l}$ from heap flow simulations for three ($R_\mathrm{d}$, $R_\mathrm{\rho}$) pairs including the example in Figure~\ref{fig1}.
For all three cases, the average segregation fluxes (data points) for the small light species (blue) and the large heavy species (red) are always equal and opposite at any particular local value of $c_\mathrm{l}$, as expected for the range of size ratios studied.\cite{prasad2017subjamming}
The curves through the data in Figure~\ref{fig2} are best fits of the cubic segregation flux model,\cite{duan2021modelling}
\begin{equation}
\phi_{\mathrm{seg},i}=d_\mathrm{s}\dot\gamma c_i [A_i+B_i(1-c_i)] (1-c_i),
\label{eq1}
\end{equation}
where $A_i$ and $B_i$ are fitting parameters dependent on both $R_\mathrm{d}$ and $R_\mathrm{\rho}$. 
For D-system segregation with $R_\mathrm{d}=1$ and $R_\mathrm{\rho}=4$ in Figure~\ref{fig2}a, light particles rise and heavy particles sink except at the extremes of concentration, where the flux is zero because only one species is present.
The direction of the  segregation is reflected by the segregation flux curves that are either positive (upward segregation for light particles) or negative (downward segregation for heavy particles) across the entire bidisperse concentration range. 
Unlike the `unidirectional' segregation case in Figure~\ref{fig2}a, size and density differences compete with each other for $R_\mathrm{d}=2$ and $R_\mathrm{\rho}=4$ in Figure~\ref{fig2}b and for $R_\mathrm{d}=1.5$ and $R_\mathrm{\rho}=4$ in Figure~\ref{fig2}c, and the segregation direction is concentration dependent. 
For $R_\mathrm{d}=2$ and $R_\mathrm{\rho}=4$  small, light particles segregate upward for small values of $c_\mathrm{l}$ and downward for large values of $c_\mathrm{l}$.  The segregation flux direction reverses at $c_\mathrm{l,eq}=0.36$ where $\phi_{\mathrm{seg},i}=0$.
For $R_\mathrm{d}=1.5$ and $R_\mathrm{\rho}=4$, a similar situation occurs except that $c_\mathrm{l,eq}$ increases to 0.47. In addition, the overall concentration averaged segregation flux is smaller for the case in Figure~\ref{fig2}c than the case in Figure~\ref{fig2}b due to the near balance between the two segregation mechanisms across the full range of concentrations.

Repeating heap flow simulations like that shown in Figure~\ref{fig1} over many ($R_\mathrm{d}$, $R_\mathrm{\rho})$ combinations results in figures analogous to those in Figure~\ref{fig2} from which $c_\mathrm{l,eq}$ can be obtained.  Using this approach, the dependence of $c_\mathrm{l,eq}$ on $R_\mathrm{d}$ and $R_\mathrm{\rho}$ was previously explored for $1 \le R_\mathrm{d}\le 2$ and $1 \le R_\mathrm{\rho}\le 4$.~\cite{duan2021modelling}
Here we extend $c_\mathrm{l,eq}$ to $1\le R_\mathrm{d}\le 3$ and $1\le R_\mathrm{\rho}\le 5$. 
The curves in Figure~\ref{fig3} are cubic fits to the interpolated data for $c_\mathrm{l,eq}$ forced to pass through ($R_\mathrm{d}=1$, $R_\mathrm{\rho}=1$), where the segregation is necessarily zero. Particles remain mixed along the curve for $c_\mathrm{l}=c_\mathrm{l,eq}$ for the corresponding $R_\mathrm{d}$ and $R_\mathrm{\rho}$.
For $R_\mathrm{d}$ and $R_\mathrm{\rho}$ combinations to the left of or above the iso-concentration curve for $c_\mathrm{l}=c_\mathrm{l,eq}$, large particles rise, while large particles sink for combinations to the right of or below the iso-concentration curve for  $c_\mathrm{l}=c_\mathrm{l,eq}$.
Along each axis are regions (colored red or blue) where the segregation is uni-directional, corresponding to cases like that shown in Figure~\ref{fig2}a.  
Size segregation dominates regardless of density in a region adjacent to the vertical axis where large particles rise even if they are slightly heavier than the small particles (red), and density segregation dominates in a narrow band adjacent to the horizontal axis where large particles sink because they are heavier than the small particles (blue).

As previously described,\citep{duan2021modelling, duan2021designing} Figure~\ref{fig3} allows an approach for the intentional design of bidisperse particle mixtures that avoid segregation.  For instance, in many industrial situations the material of each particle species is specified, thereby fixing the density ratio, and the concentration of each species is fixed based on the product requirements. However, the species sizes can be altered by one of several standard processes such as agglomeration or grinding.  Thus, it is possible to specify a size ratio for a given density ratio and relative concentration of species that minimizes segregation.  For example, suppose that a 20:80 species concentration mixture is required and the density ratio of the two species is $R_\mathrm{\rho}=2.5$.  Starting at $R_\mathrm{\rho}=2.5$ and reading upward to the $c_\mathrm{l,eq}=0.2$ contour in Figure~\ref{fig3} indicates that a size ratio of $R_\mathrm{d}=1.6$ should result in a non-segregating mixture.  
In the remainder of this paper, we explore this approach to designing minimally-segregating bidisperse particle systems.

 Before continuing with the approach of designing minimally-segregating particle mixtures  by using the appropriate combination of  $R_\mathrm{d}$, $R_\mathrm{\rho}$, and $c_\mathrm{l,eq}$, we consider briefly an alternative approach that ignores $c_\mathrm{l,eq}$ and is therefore more general.  Returning to Figure~\ref{fig2}, it is evident that the magnitude of the segregation flux depends on size and density ratios in addition to the mixture concentration.  For instance, the segregation flux across all values of $c_\mathrm{l}$ for $R_\mathrm{d}=1.5$ and $R_\mathrm{\rho}=4$ in Figure~\ref{fig2}c is generally less than that for $R_\mathrm{d}=2$ and $R_\mathrm{\rho}=4$ in Figure~\ref{fig2}b.  Hence, a system with $R_\mathrm{d}=1.5$ and $R_\mathrm{\rho}=4$ would  tend to segregate less, regardless of mixture concentration, than a system with $R_\mathrm{d}=2$ and $R_\mathrm{\rho}=4$.  Using data across the full range of $R_\mathrm{d}$ and $R_\mathrm{\rho}$ considered here, we quantify this average tendency toward segregation at a particular combination of $R_\mathrm{d}$ and $R_\mathrm{\rho}$ as the square of the dimensionless segregation flux averaged over all possible concentrations, 
 \begin{equation}
 {\bar\Phi}^2=\int_0^1 (\phi_{\mathrm{seg},\mathrm{l}}/\dot\gamma d_\mathrm{s})^2 d c_\mathrm{l},
 \end{equation}
  which is plotted in Figure~\ref{fig4} as a function of size and density ratios. 
The grayscale reflects the sensitivity of the system to segregation across all possible concentrations $c_\mathrm{l}$. 
Combinations of $R_\mathrm{d}$ and $R_\mathrm{\rho}$ in brighter areas of Figure~\ref{fig4} tend to segregate less across all mixture concentrations, while combinations in darker areas tend to segregate more.  
The least  segregation occurs for combinations of $R_\mathrm{d}$ and $R_\mathrm{\rho}$ along the dashed curve, which is calculated as $\partial{\bar\Phi}^2/\partial R_\mathrm{d}=0$.\footnote[3]{For application, the curve can be approximated as a linear function $R_\mathrm{d}=0.163(R_\mathrm{\rho}-1)+1$.}
Interestingly, this curve for the least segregation corresponds closely to the $c_\mathrm{l,eq}=0.5$ curve from Figure~\ref{fig3} which is reproduced in Figure~\ref{fig4} as the solid curve.  
This is because the lowest flux magnitudes tend to occur when $c_\mathrm{l,eq}$ is around 0.5 as is evident in comparing Figures~\ref{fig2}b and \ref{fig2}c.

Figure~\ref{fig4} offers an alternative approach to the one based on  Figure~\ref{fig3} where $\bar c_\mathrm{l}$ is set to $c_\mathrm{l,eq}$ to prevent segregation.  This alternative is simply to ignore the mixture concentration and operate in the light colored region of Figure~\ref{fig4} along the dashed curve for the minimum value of ${\bar\Phi}^2$.  In this case, for a given density ratio for the two particle species, the size ratio would be adjusted to fall on the minimum segregation curve in Figure~\ref{fig4}.  For instance, for the preceding example with a density ratio $R_\mathrm{\rho}=2.5$, the size ratio should be set to $R_\mathrm{d}\approx 1.3$ to minimize segregation across all possible concentrations.  While this approach will result in more segregation when $c_\mathrm{l}\neq c_\mathrm{l,eq}$, it may be easier to implement in some industrial situations.  
However, since the approach in Figure~\ref{fig3} using $c_\mathrm{l,eq}$ is more precise in avoiding segregation than the alternative approach based on Figure~\ref{fig4}, we focus on the former in the remainder of this paper.

\section{Validating predictions for non-segregating mixtures}

The phase diagram in Figure~\ref{fig3} is based on the local segregation flux, and the equilibrium concentration $c_\mathrm{l,eq}$ is inherently a local variable at each point within the thin flowing layer typical of many gravity-driven granular flows.
However, the requirement that $c_\mathrm{l}=c_\mathrm{l,eq}$ everywhere in the flow to prevent segregation is unlikely to be realized exactly in real systems due to the stochastic nature of granular flows which drives local variations in the mixture concentration, as well as to deviations in the actual velocity profile from that of the bounded heap for which the $c_\mathrm{l,eq}$ curves in Figure~\ref{fig3} were obtained.  
Furthermore, the averaging approach used to obtain Figure~\ref{fig3} and Equation~\ref{eq1} ignores the potential influence of absolute pressure~\cite{fry2018effect} and pressure variation in combination with the effects of the shear rate gradient\cite{jing2021unified} on segregation.
Hence, the question is if a uniformly mixed system with a global concentration of $\bar c_\mathrm{l}= c_\mathrm{l,eq}$ remains mixed or locally segregates across different flow geometries and initial conditions.  In this section, we validate and demonstrate the potential for designing minimally-segregating granular mixtures using the approach of specifying $R_\mathrm{d}$, $R_\mathrm{\rho}$, and $c_\mathrm{l,eq}$ (Figure~\ref{fig3}) by quantifying and comparing the segregation of particles near the equilibrium conditions shown in Figure~\ref{fig3} predicted by Eq.~\ref{eq1} in bounded heaps and rotating tumblers.  The goal is to determine how the \textit{local} propensity for mixing or segregation shown in Figure~\ref{fig3} affects the \textit{global} segregation across the entire flow domain.

\subsection{Bounded heap flow}

\begin{figure}[t]
\centering
\includegraphics[width=\columnwidth]{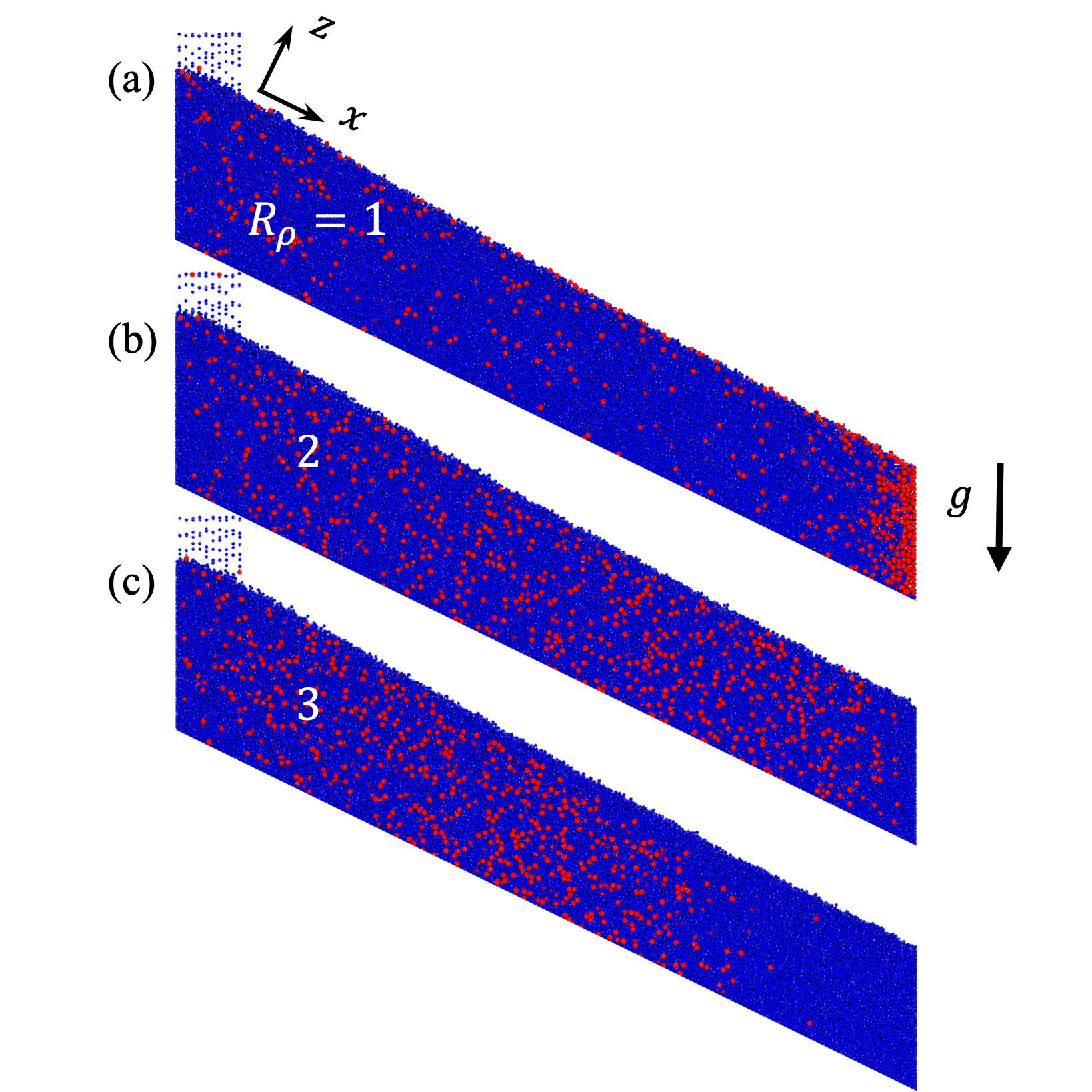}
\caption{Heap flow segregation examples from DEM simulations for feed concentration of large red particles $ \bar c_\mathrm{l}=0.1$ and size ratio $R_\mathrm{d}=1.5$ showing reversal in large particle segregation direction for different density ratios $R_\mathrm{\rho}$. Other conditions are as in Figure~\ref{fig1}.
}
\label{fig5}
\end{figure}

We consider first the global segregation of particle mixtures prepared at the equilibrium concentration (i.e., $\bar c_\mathrm{l}=c_\mathrm{l,eq}$) and fed onto a bounded heap.  Figure~\ref{fig5} shows the segregation patterns resulting from a feed concentration of $\bar c_\mathrm{l}=0.1$, $R_\mathrm{d}=1.5$, and three different values of $R_\mathrm{\rho}$.  For $R_\mathrm{\rho}=1$ large red particles segregate upward to the free surface of the flowing layer and deposit downstream near the endwall, indicating that size-based percolation dominates. Particles remain relatively mixed over most of the heap for $R_\mathrm{\rho}=2$, as the two segregation mechanisms nearly balance. Segregation reverses for $R_\mathrm{\rho}=3$ with small light particles (blue) depositing at the downstream end of the heap as buoyancy dominates over percolation. 

\begin{figure}[t]
\centering
\includegraphics[width=\columnwidth]{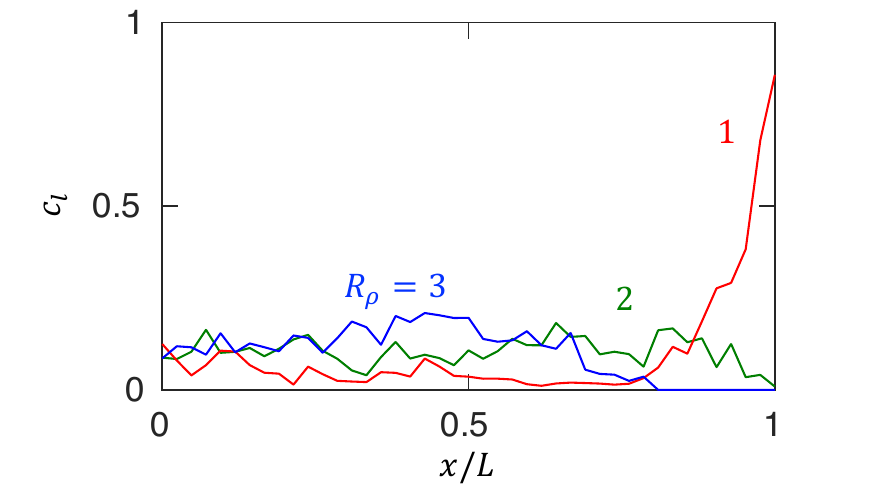}
\caption{
Streamwise concentration profiles of large particles, $c_\mathrm{l}$, deposited on the heap for the three different $R_\mathrm{\rho}$ values in Figure~\ref{fig4} with $\bar c_\mathrm{l}=0.1$ and $R_\mathrm{d}=1.5$.
}
\label{fig6}
\end{figure}

To quantify these results, Figure~\ref{fig6} plots streamwise $c_\mathrm{l}$ profiles in the deposited heap for the three cases in Figure~\ref{fig5}.  
For $R_\mathrm{\rho}=1$ (red curve), the concentration of large particles is highest near $x/L\approx1$ as they deposit on the downstream portion of the heap along with a few small particles, consistent with Figure~\ref{fig5}a.
The situation reverses for $R_\mathrm{\rho}=3$ (blue curve), where a higher fraction of large particles deposit on the upstream portion of the heap, and pure small particles ($c_\mathrm{l}=0$) deposit on the downstream portion, as shown in Figure~\ref{fig5}c.
For the intermediate density ratio where size and density effects nearly balance, large particles deposit almost uniformly on the heap and the streamwise variation in $c_\mathrm{l}$ (green curve) is the smallest of the three cases.

The degree and direction of segregation along the length of the heap is calculated using the signed and scaled standard deviation of the large particle concentration,
\begin{equation}
{\hat \sigma_l}=\mathrm{sgn}(c_{\mathrm{l},N}-\langle c_\mathrm{l} \rangle) \frac{\sqrt{ {\sum_{k=1}^N (c_{\mathrm{l},k}-\langle c_\mathrm{l}\rangle )^2}/{N}} } { {\langle c_l\rangle (1-\langle c_l\rangle)}},
\label{xl}
\end{equation}
where $N=50$ is the number of uniform width bins for calculating $c_\mathrm{l}$ at different streamwise positions, $c_{\mathrm{l},k}$ is the local depth-averaged volume concentration for particles deposited on the heap below the flowing layer in bin $k$, and $\langle c_\mathrm{l}\rangle$ is the average value of $c_\mathrm{l}$ for $0 \le x \le L$. Note that $\langle c_\mathrm{l}\rangle$ differs slightly from $\bar{c}_\mathrm{l}$ because it excludes the portion of the heap below the feed zone (see Figure 1).  The value $\hat{\sigma}_\mathrm{l}$ is essentially the standard deviation of $c_\mathrm{l}$, but with two additional multiplicative terms.  The sign function, $\mathrm{sgn}(c_{\mathrm{l},N}-\langle c_\mathrm{l}\rangle)$, indicates the large particle segregation direction: a value of $-1$ indicates downward segregation resulting in $c_\mathrm{l}>\langle{c}_\mathrm{l}\rangle$  on the upstream portion of the heap and $c_{\mathrm{l},N}=c_\mathrm{l}(x=L)<\langle{c}_\mathrm{l}\rangle$ at the downstream end; a value of $+1$ corresponds to upward segregation resulting in $c_\mathrm{l}<\langle{c}_\mathrm{l}\rangle$ on the upstream portion of the heap and $c_{\mathrm{l},N}>\langle{c}_\mathrm{l}\rangle$ at the downstream end.  The $\langle c_l\rangle (1-\langle c_l\rangle)$ term in the denominator normalizes the measured standard deviation by that for perfect segregation. A fully mixed, non-segregating case with large particles depositing uniformly along the surface of the heap has $\hat \sigma_\mathrm{l}=0$, while 
for complete segregation, $\hat \sigma_\mathrm{l}=-1$ for sinking large particles and $\hat \sigma_\mathrm{l}=1$ for rising large particles.

\begin{figure}[t]
\centering
\includegraphics[width=\columnwidth]{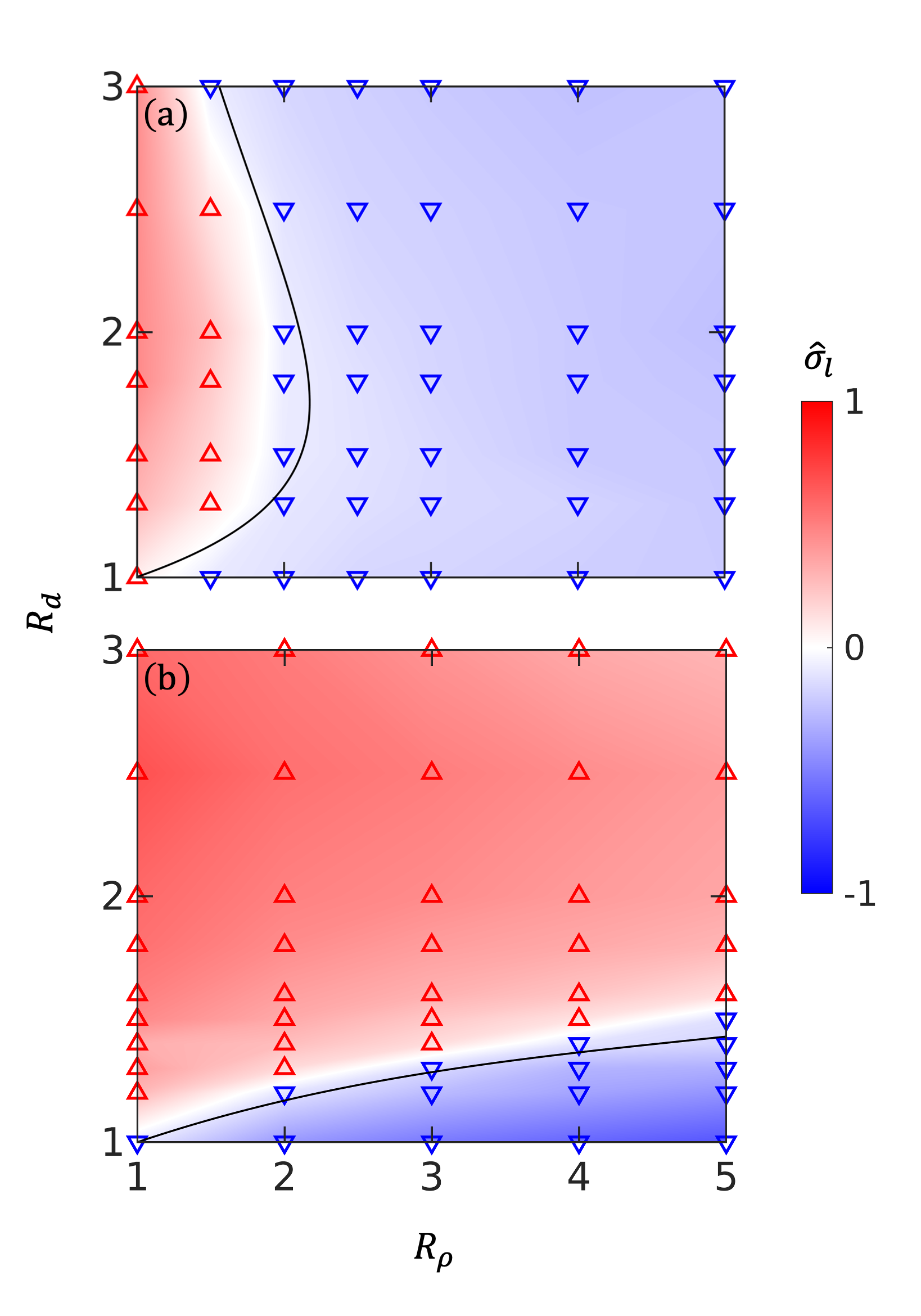}
\caption{Scaled deviation of large particle concentration, $\hat \sigma_l$, (color contours) for feed concentration $\bar c_\mathrm{l}$ of (a) 0.1 and (b) 0.6. Black curves show predicted non-segregating $R_\mathrm{d}$ and $R_\mathrm{\rho}$ values for those feed concentrations; white regions correspond to particles remaining mixed in heap simulations.
Upward red triangles indicate large particles rise; downward blue triangles indicate large particles sink.}
\label{fig7}
\end{figure}

Figure~\ref{fig7} demonstrates how $\hat \sigma_l$ captures the rise-sink transition in heap flow simulations (data points) for 49 ($R_\mathrm{d}$, $R_\mathrm{\rho}$) combinations with $\bar c_\mathrm{l}=0.1$  (Figure~\ref{fig7}a) and 50 ($R_\mathrm{d}$, $R_\mathrm{\rho}$) combinations with $\bar c_\mathrm{l}=0.6$ (Figure~\ref{fig7}b). 
Red and blue shading based on  interpolating these data corresponds to the degree to which $\hat \sigma_l$ deviates from the perfectly mixed state value of $\hat \sigma_l=0$ with white corresponding to particles remaining mixed.
The predicted equilibrium curve for the local value of $c_\mathrm{l}$ from Figure~\ref{fig3}, shown by the black curve,  corresponds closely to the white region, indicating that particles remain mixed at that feed concentration.  In other words, the $c_\mathrm{l,eq}$ curve in the $R_\mathrm{d}R_\rho$-plane along which particles are predicted to locally remain mixed from Figure~\ref{fig3} (black curve) corresponds closely to ($R_\mathrm{d}$,$R_\mathrm{\rho}$) pairs along which particles remain globally mixed for that feed concentration under heap flow (white region where $\hat \sigma_l=0$).
The small deviation of the black curve from the white region could result from many factors including the resolution of data points in the ($R_\mathrm{d}$,$R_\mathrm{\rho}$) space and fluctuations in the local concentration similar to those observed for the weakly segregating case (green curve) shown in Figure~\ref{fig6}.

\begin{figure}[t]
\centering
\includegraphics[width=\columnwidth]{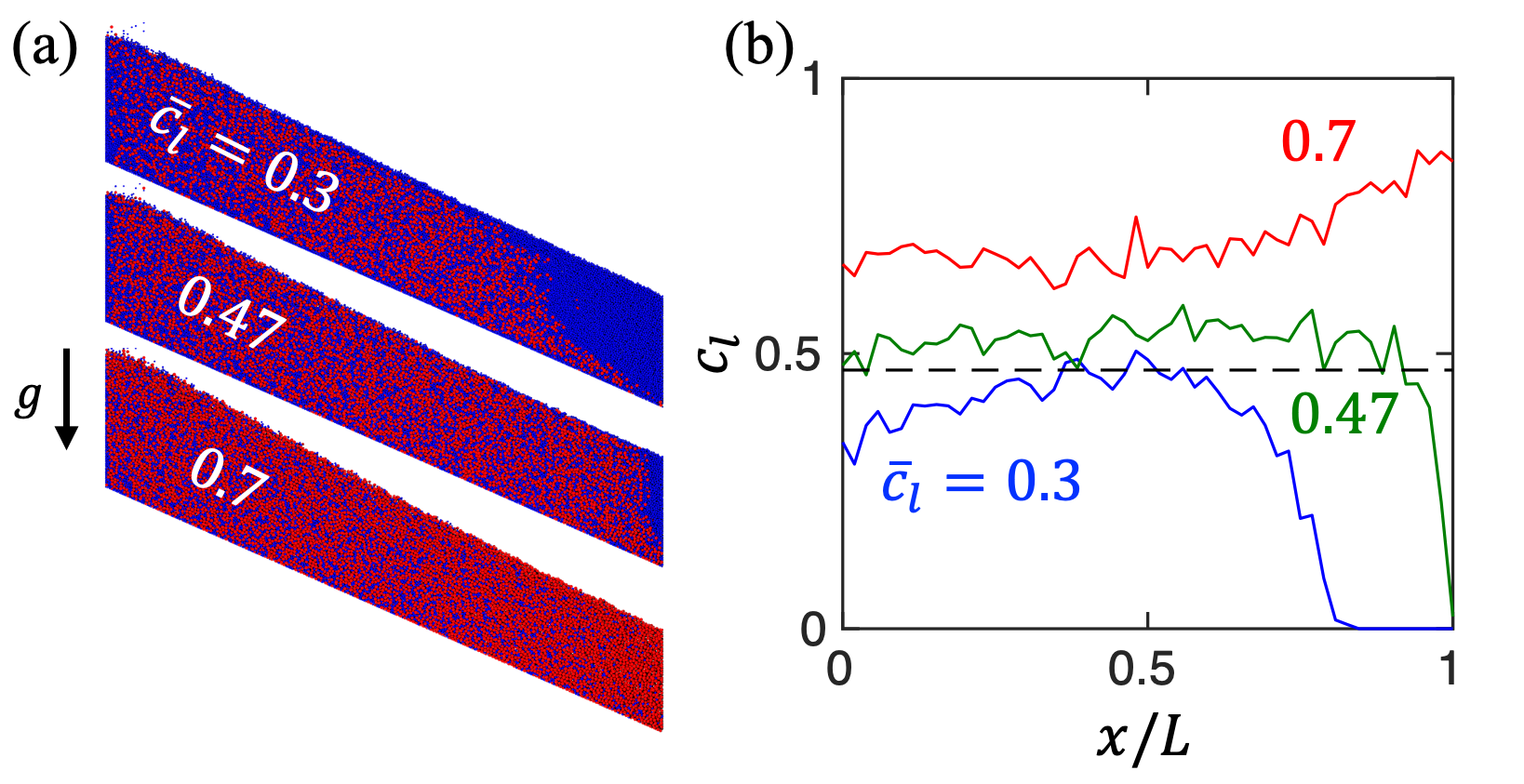}
\caption{
 (a) DEM simulation images and (b) large particle concentration vs.\ streamwise position for heap flow segregation with $R_\mathrm{d}=1.5$, $R_\mathrm{\rho}=4$, and different values of large particle feed concentration $\bar c_\mathrm{l}$. Large heavy particles (\textcolor{black}{red}) sink while small light particles (\textcolor{black}{blue}) rise for $\bar c_\mathrm{l}=0.3<c_\mathrm{l,eq}$, as buoyancy overcomes percolation. In contrast, for $\bar c_\mathrm{l}=0.7>c_\mathrm{l,eq}$ segregation is reversed and percolation dominates over buoyancy.  Particles remain relatively mixed for $\bar c_\mathrm{l}= c_\mathrm{l,eq}=0.47$ (horizontal dashed line).
}
\label{fig_heap}
\end{figure}

Particle mixtures with ($R_\mathrm{d},R_\mathrm{\rho}$) off the equilibrium curve in Figure~\ref{fig3} segregate, as expected. 
Conditions below and to the right of the $ c_\mathrm{l,eq}=\bar c_\mathrm{l}$  equilibrium curve in the $R_\mathrm{d} R_\mathrm{\rho}$-plane correspond to smaller size ratios and larger density ratios, indicating that large particles sink for these conditions; conditions above and to the left of the equilibrium curve correspond to larger size ratios and smaller density ratios, indicating that large particles rise.

An equivalent alternate description is that if the feed concentration $\bar c_\mathrm{l}$ for a particular ($R_\mathrm{d},R_\mathrm{\rho}$) pair is above the equilibrium value in Figure~\ref{fig3} ($\bar c_\mathrm{l}>c_\mathrm{l,eq}$), then the feed concentration is too high to maintain equilibrium, and the large particles will rise resulting in $\hat \sigma_l>0$.  
If the feed concentration $\bar c_\mathrm{l}$ for the ($R_\mathrm{d},R_\mathrm{\rho}$) pair is below the  equilibrium value in Figure~\ref{fig3} ($\bar c_\mathrm{l}<c_\mathrm{l,eq}$), then the feed concentration is too low to maintain equilibrium, and large particles will sink resulting in $\hat \sigma_l<0$. 
This is evident in the segregation examples with $R_\mathrm{d}=1.5$, $R_\mathrm{\rho}=4$, and $c_\mathrm{l,eq}=0.47$ shown in Figure~\ref{fig_heap}. Large heavy  particles (red) sink for $\bar c_\mathrm{l}=0.3<c_\mathrm{l,eq}$, depositing on the upstream portion of the heap, whereas for $\bar c_\mathrm{l}=0.7>c_\mathrm{l,eq}$ large heavy particles (red) rise, depositing at a higher concentration on the downstream portion of the heap. For $\bar c_\mathrm{l}=c_\mathrm{l,eq}$ particles remain well mixed except near the downstream bounding endwall where $c_\mathrm{l}$ decreases somewhat, potentially due to a pressure dependence of the equilibrium concentration that is unaccounted for in Equation~(\ref{eq1}) and the predictions of Figure~\ref{fig3}, as discussed shortly with regard to the rotating tumbler results.

The implication of these results for heap flow is that the local propensity for mixing or segregation as predicted in Figure~\ref{fig3} also reflects the global propensity for mixing or segregation, at least to a first order.  The consequence is that if granular material with $\bar c_\mathrm{l}=c_\mathrm{l,eq}$ is mixed at the feed, it remains mixed as it deposits on the heap. 
In addition, the facts that particles remain relatively mixed for ($R_\mathrm{d}$,$R_\mathrm{\rho}$) pairs near the equilibrium curve as shown in Figure~\ref{fig7}, and that $\hat \sigma_\mathrm{l}$ varies smoothly across the ($R_\mathrm{d}$,$R_\mathrm{\rho}$) space in Figure~\ref{fig7} indicate that slight deviation of $\bar c_\mathrm{l}$ from $c_\mathrm{l,eq}$ or small fluctuations in $c_\mathrm{l}$ do not affect the overall propensity for particles to remain mixed in heap flows for appropriate values of $R_\mathrm{d}$ and $R_\mathrm{\rho}$.

\subsection{Rotating tumbler flow}

\begin{figure}[t]
\centering
\includegraphics[width=\columnwidth]{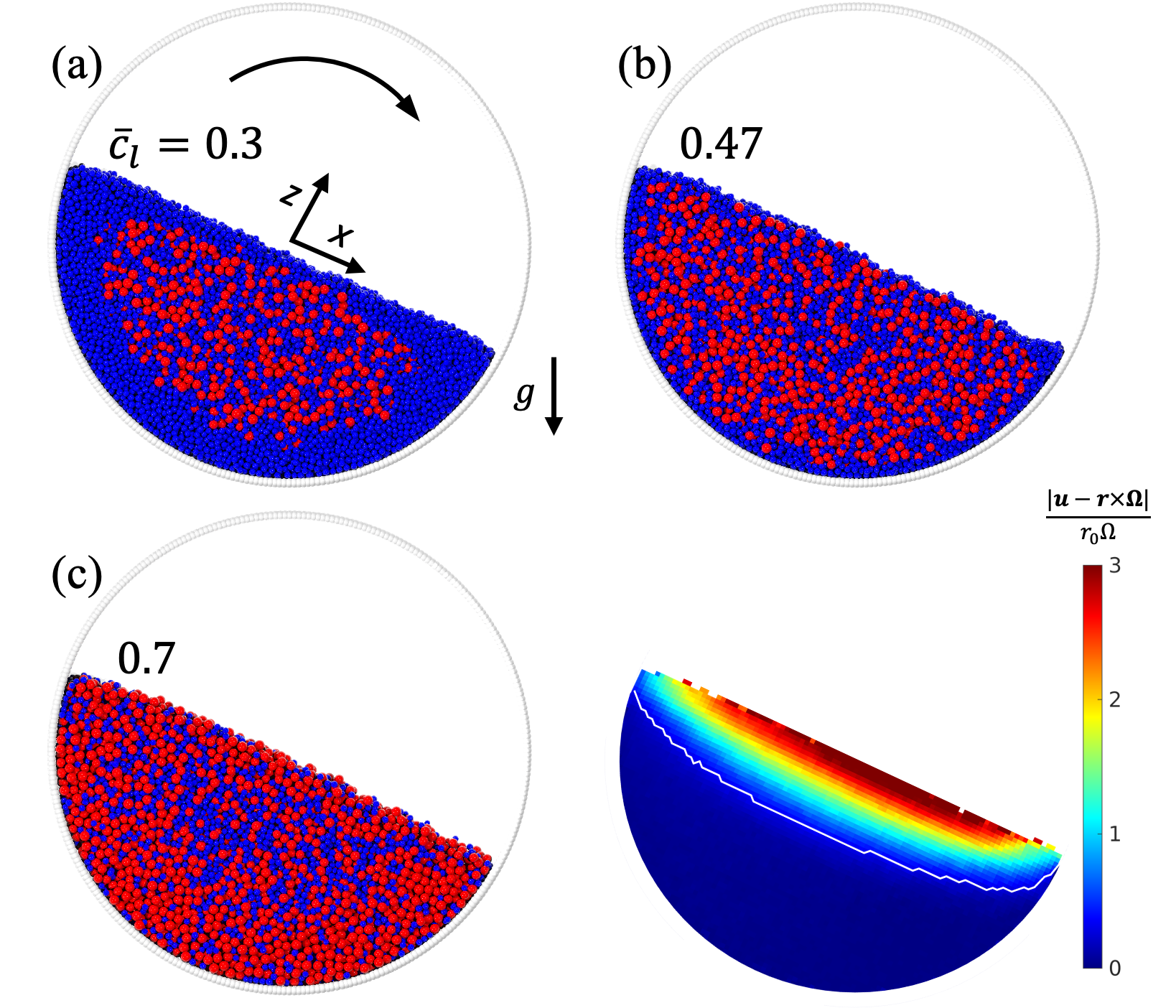}
\caption{Rotating tumbler DEM simulation setup and segregation examples in steady state for $R_\mathrm{d}=1.5$, $R_\mathrm{\rho}=4$, and global large particle concentration $\bar c_\mathrm{l}$ of (a) 0.3, (b) 0.47, and (c) 0.7 after 7 rotations, or 42\,s, starting from a fully mixed initial condition.
Large particles are \textcolor{black}{red} ($d_\mathrm{l}=3\,$mm, $\rho_\mathrm{l}=4\,$g/cm$^3$) and small particles are \textcolor{black}{blue} ($d_\mathrm{s}=2\,$mm, $\rho_\mathrm{s}=1\,$g/cm$^3$). 
(d) Non-dimensional speed in rotating reference frame, $| \bm u-\bm r\times \bm\Omega|/r_0\Omega$, for all particles in case (b) averaged over 10\,s. Solid curve shows  where the streamwise velocity in the lab frame is zero, i.e., $u_x(x,z)=0$.
 }
\label{fig9}
\end{figure}

In the previous section, we demonstrated the application of our approach for designing minimally-segregating mixtures to heap flows.  Since the data for the local equilibrium concentrations in Figure~\ref{fig3} upon which the approach is based come from segregation in heap flows, its effectiveness is, perhaps, not surprising.  The question now is if the equilibrium concentrations obtained from heap flows can be used to identify minimally-segregating mixtures in a flow geometry where the velocity field and boundary conditions differ.  Here, we apply the approach to rotating tumbler flow.\cite{metcalfe1996pattern,jain2005regimes,pereira2014segregation,umbanhowar2019modeling}

Tumblers are used to coat, crush, and mix particles. 
Unlike bounded heaps where segregation takes place during the short period of time before the particles deposit onto the fixed bed,  segregation in rotating tumblers is an ongoing process as particles repeatedly flow down the slope, enter solid-body rotation in the downstream half of the flowing layer, and then re-enter the upper half of the flowing layer after solid body rotation.
There is an initial transient as initially mixed particles segregate in the flowing layer and then deposit in a segregated pattern in the first half-rotation, followed by enhanced segregation as the particles repeatedly flow down the surface.  
A steady segregated pattern is established after only a few tumbler rotations\cite{nityanand1986analysis,cantelaube1995radial,schlick2015granular}  in which segregated particles in the fixed bed enter the flowing layer, flow down the surface, and maintain their segregated pattern upon re-entering the fixed bed.

Here we test the same bidisperse particle mixtures as used in the heap flow simulations described in the previous section, noting that $\bar{c}_\mathrm{l}$ here refers to the overall large particle concentration in the tumbler rather than the feed concentration, as was the case for the heap.
The half-full tumbler is $r_0=7.5~$cm in radius ($r_0/d_\mathrm{l}=25$) and 1.5 cm in length with periodic boundaries in the axial direction to avoid endwall effects.
The cylindrical wall of the tumbler is formed from 3\,mm particles overlapping by 1.5\,mm to reduce slip between the particle bed and the wall.  
The tumbler rotates clockwise at $\Omega=10$~rev/min. The Froude number is $F_r=\Omega^2 r_0/g=0.0084$, which corresponds to the flat surface continuous flow regime.\cite{mellmann2001transverse}

Figure~\ref{fig9} shows examples of the steady segregation pattern after 7 rotations ($t\Omega=7$) of the tumbler for the same particle mixtures used in the bounded heap examples  in Figure~\ref{fig_heap} starting with a well-mixed initial condition.
The equilibrium concentration of large particles for $R_\mathrm{d}=1.5$ and $R_\mathrm{\rho}=4$ is $c_\mathrm{l,eq}=0.47$ according to  Figures~\ref{fig2}c and~\ref{fig3}.
Similar to large heavy particles depositing in the upstream portion of the heap for $\bar c_\mathrm{l}<c_\mathrm{l,eq}$, large heavy particles sink in the flowing layer and segregate to the core of the tumbler bed for $\bar c_\mathrm{l}=0.3$ in Figure~\ref{fig9}a.
Figure~\ref{fig9}b shows that for $\bar c_\mathrm{l}= c_\mathrm{l,eq}=0.47$ particles remain relatively mixed, as the two segregation mechanisms nearly balance. 
Segregation is reversed at $\bar c_\mathrm{l}=0.7>c_\mathrm{l,eq}$ in Figure~\ref{fig9}c as percolation dominates over buoyancy, i.e., large heavy particles segregate to the tumbler periphery analogous to their deposition in the downstream portion of the heap.

For reference, Figure~\ref{fig9}d shows the non-dimensional particle speed in the rotating reference frame of the tumbler for $\bar c_\mathrm{l}=0.47$, i.e., $| \bm u- \bm r\times \bm\Omega |/r_0\Omega$ where $\bm u$ is measured in the lab frame.
The speed of the particles is largest at the midpoint of the free surface of the flowing layer and decreases with depth.
Segregation takes place in the flowing surface layer due to shear; below this layer particles are in near solid-body rotation with the tumbler and do not segregate.
The elapsed time between particle passes through the flowing layer (i.e. the solid body rotation residence time) is slightly less than half of the tumbler rotation period, and the time a particle typically spends in the flowing layer is one order of magnitude less than the tumbler rotation period.\cite{zaman2013slow}
Similar velocity fields are observed for the other two values of $\bar c_\mathrm{l}$.

\begin{figure}[t]
\centering
\includegraphics[width=\columnwidth]{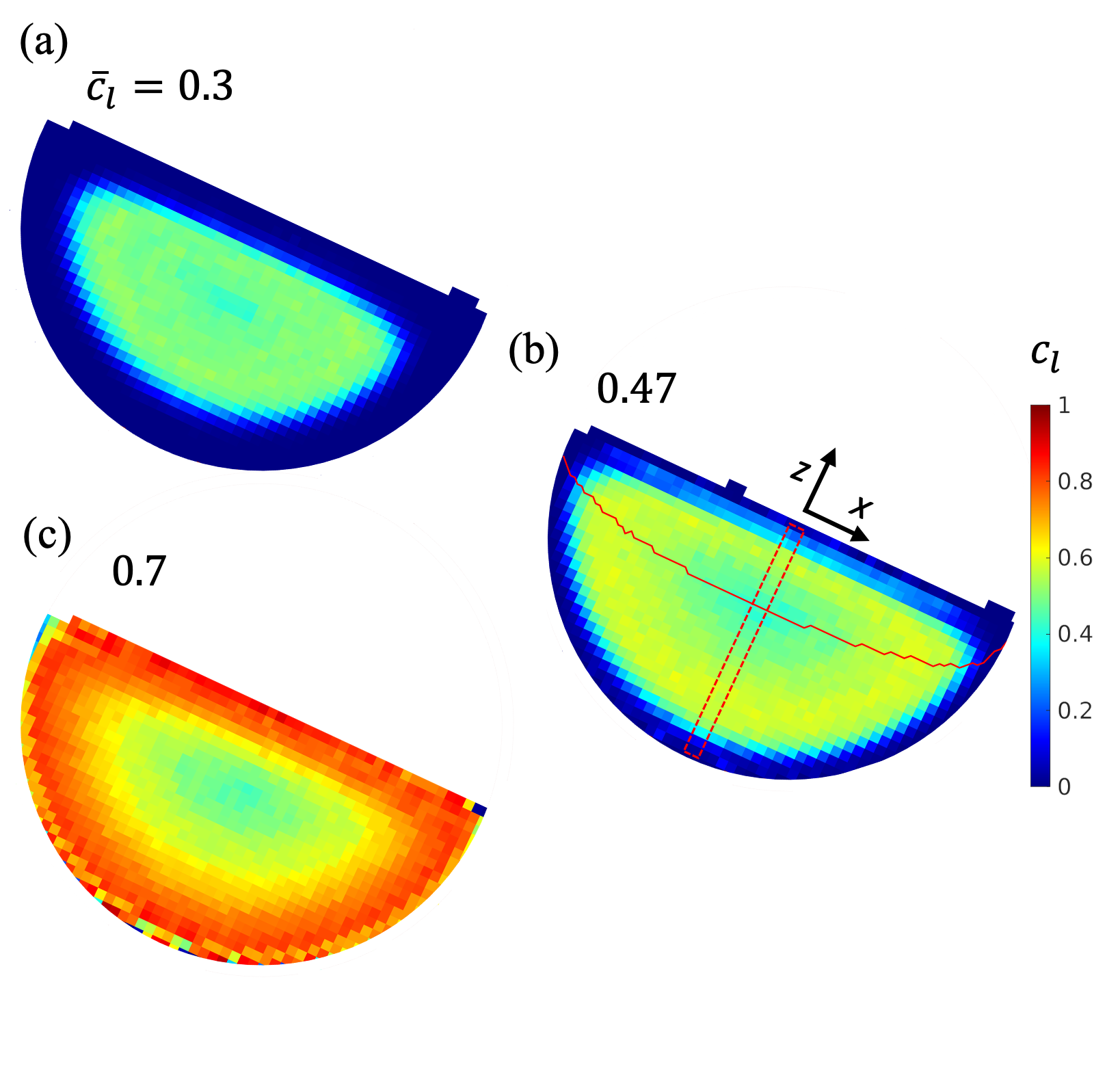}
\caption{
Steady-state large-particle concentration $c_\mathrm{l}$ averaged over 10\,s after 7 rotations (42\,s) in the particle-filled portion of the tumbler rotating at 10\,rpm for $R_\mathrm{d}=1.5$, $R_\mathrm{\rho}=4$, and $\bar c_\mathrm{l}$ equal to (a) 0.3, (b) 0.47 and (c) 0.7. 
Solid curve in (b) shows the bottom of the flowing layer where the streamwise velocity in the lab frame is zero, i.e., $u_x(x,z)=0$. Rectangular box in (b) is the region used for $c_\mathrm{l}$ profiles.
}
\label{fig10}
\end{figure}

\begin{figure}[t]
\centering
\includegraphics[width=\columnwidth]{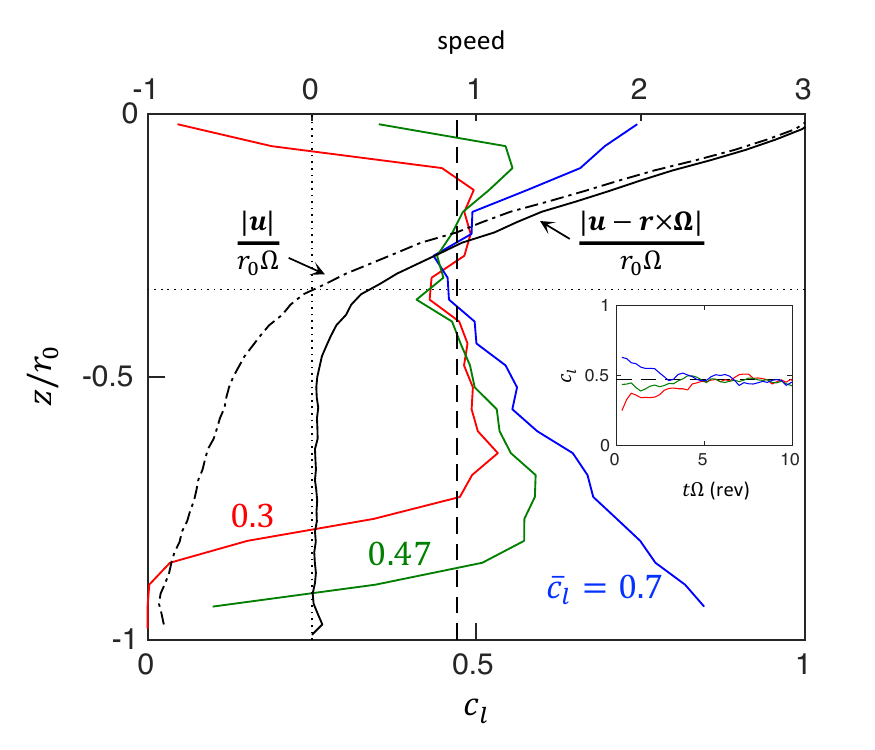}
\caption{ 
Depth profile of scaled  particle speed in the lab frame (dash-dot black curve) and in the rotating tumbler frame (solid black curve) along a radius normal to the free surface and averaged across the region indicated by the dashed rectangle in Figure~\ref{fig10}b.
Dotted horizontal  line at $z/r_0= -0.3$ indicates the approximate bottom of the flowing layer where $u_x (x=0,z=-\delta)\approx 0$ and which also corresponds to the core center. 
Colored curves are radial profiles of average large particle concentration normal to the free surface, as indicated by the dashed rectangle in Figure~\ref{fig10}b, in steady state for three different $\bar c_\mathrm{l}$ values. 
At the core's approximate center ($z/r_0=-0.3$, dotted horizontal line), $c_\mathrm{l} \approx c_\mathrm{l,eq}$ (dashed vertical line) even when $\bar c_\mathrm{l} \neq c_\mathrm{l,eq}$.  
Inset: $c_\mathrm{l}$ at the core center ($z/r_0=-0.3$) tends toward the equilibrium concentration, $c_\mathrm{l,eq}$ with increasing number of rotations regardless of the overall concentration, $\bar c_\mathrm{l}$.
 }
\label{fig11}
\end{figure}

The  core region of the tumbler bed is typically thought of as the inner part of the bed of particles where small (S-system) or heavy (D-system) particles segregate, which is surrounded by particles of the other species (large or light particles, respectively). It includes particles that are both in the flowing layer and in the fixed bed, i.e.\ in solid body rotation.  
This is evident comparing the lower bound of the flowing layer (white contour in Figure~\ref{fig9}d) with the extent of large particle rich core region in Figure~\ref{fig9}a. 
The core region spans the lower portion of the flowing layer and the inner portion of the solid body rotation region.
This is most easily understood in terms of steady-state pathlines, which are, by definition, circular arcs in the portion of the bed in solid body rotation and generally in the streamwise direction, though slightly curved, in the flowing layer.\cite{khakhar1997radial,jain2005regimes}  
For steady-state segregation, there is a one-to-one correspondence between the particle species distribution in the fixed bed and that in the flowing layer.  Hence, the core region is reflected in both the flowing layer and the bed in solid body rotation with the core center (rotation center of the particles) at the boundary between them along a radial line at $x=0$. 
Here, we define the approximate center of the tumbler core as the point where the velocity field is zero (white contour in Figure~\ref{fig9}d) at the midpoint of the flowing layer ($x=0$). 
For all cases studied here, this location is well approximated by a constant value of $z/r_0=-0.3$.

The local concentration of large particles in the three rotating tumbler cases in Figure~\ref{fig9} can be calculated from spatial and temporal averages of simulation data after the system reaches steady-state. 
Figure~\ref{fig10} shows the steady state concentration of large particles. 
In the core of the tumbler bed, away from the flowing surface layer and cylindrical tumbler wall, $c_\mathrm{l}$ is nearly the same for all three cases, while the concentration surrounding the core and outward to the periphery differs significantly between the three cases.
For $\bar c_\mathrm{l}=0.3<c_\mathrm{l,eq}$, $c_\mathrm{l}$ at the tumbler periphery is effectively zero, consistent with Figure~\ref{fig9}a where small light  particles (blue) segregate to the tumbler periphery.
In contrast, for $\bar c_\mathrm{l}=0.7>c_\mathrm{l,eq}$ the segregation direction is reversed, and $c_\mathrm{l}$ increases approaching the tumbler periphery, consistent with an excess of large heavy particles (red) at the periphery in Figure~\ref{fig9}c.

The relationship between the mixture concentration and the flowing layer thickness is depicted on a more quantitative basis in Figure~\ref{fig11}.
This figure shows the depth profile of the large particle concentration, $c_\mathrm{l}$, along a radius normal to the free surface and averaged across the narrow region within the rectangle in Figure~\ref{fig10}b (indicated along the lower horizontal axis), as well as the scaled particle speed in the lab frame and in the rotating tumbler frame (indicated along the upper horizontal axis). 
Concentration data are excluded for locations where the volume fraction is below 0.3 [corresponding to near the free surface ($z/r_0\approx 0$) and near the tumbler wall ($z/r_0\approx -1$) due to the overlap of the rectangular averaging regions with the circular boundary of the tumbler].

The $c_\mathrm{l}$ concentration profiles for the three cases in Figure~\ref{fig10} have a concentration in the core, corresponding to $-0.5\lessapprox z/r_0 \lessapprox -0.2$, that is approximately equal to the equilibrium concentration, $c_\mathrm{l,eq}$, as shown in Figure~\ref{fig11}.
In particular, the profiles for all three cases overlap and have a minimum value of $c_\mathrm{l} \approx c_\mathrm{l,eq}$ (dashed vertical line) at $z/r_0 \approx -0.3$ (horizontal dotted line), the approximate radial position of the core center.
Thus, the center of the core and the region around it is mixed at the equilibrium concentration, $c_\mathrm{l,eq}$, while the particle concentration in the periphery adjusts to a concentration necessary to accommodate the remaining particles, whether they are small particles (Figure~\ref{fig10}a) or large particles (Figure~\ref{fig10}c).

Keeping in mind that the particles are initially fully mixed, the question arises about how long it takes for the core to reach the equilibrium concentration.
To answer this, the inset in Figure~\ref{fig11} shows time series of local concentration at the core center  (i.e., $c_\mathrm{l}$ at $z/r_0=-0.3$).  
The concentration $c_\mathrm{l}$ ($z/r_0=-0.3$) starts at approximately the initial concentration $\bar c_\mathrm{l}$ in each case, and the three curves converge to the equilibrium concentration within five rotations ($t\Omega=5$, or 30\,s).  
This is also about the time necessary for the global segregation pattern to reach steady state.

From the concentration profile for $\bar{c}_\mathrm{l}=c_\mathrm{l,eq}=0.47$ in Figure~\ref{fig11}, it is evident that the mixing is imperfect at the equilibrium condition.  Nevertheless, $c_\mathrm{l}$ is within a relatively narrow range $0.44 \le c_\mathrm{l} \le 0.58$ except near the free surface, $z/r_0 \lessapprox -0.1$, and near the tumbler wall, $z/r_0 \gtrapprox -0.9$, where it decreases. 
Note that to a first approximation, the concentration profile in the solid body rotation portion of the particles is simply a stretched reflection of the concentration profile in the flowing layer. This occurs because particles deposit on the steady-state pathlines described earlier.
Hence, these decreased concentrations at the surface and the tumbler wall are essentially mirror images of one another.
The reduced large particle concentration near the tumbler wall has been noted previously \citep{pereira2014segregation,thomas2018evidence} but not explained, to our knowledge.  
The deviation of  $c_\mathrm{l}$ from $c_\mathrm{l,eq}=0.47$ at the top of the flowing layer and near the cylindrical wall may be due to the effect of depth-varying lithostatic pressure, which is  unaccounted for in the equilibrium concentration results in Figure~\ref{fig3}.
An extended discussion of this deviation in $c_\mathrm{l}$ is included in Sec.~\ref{discussion}.

\begin{figure}[t]
\centering
\includegraphics[width=\columnwidth]{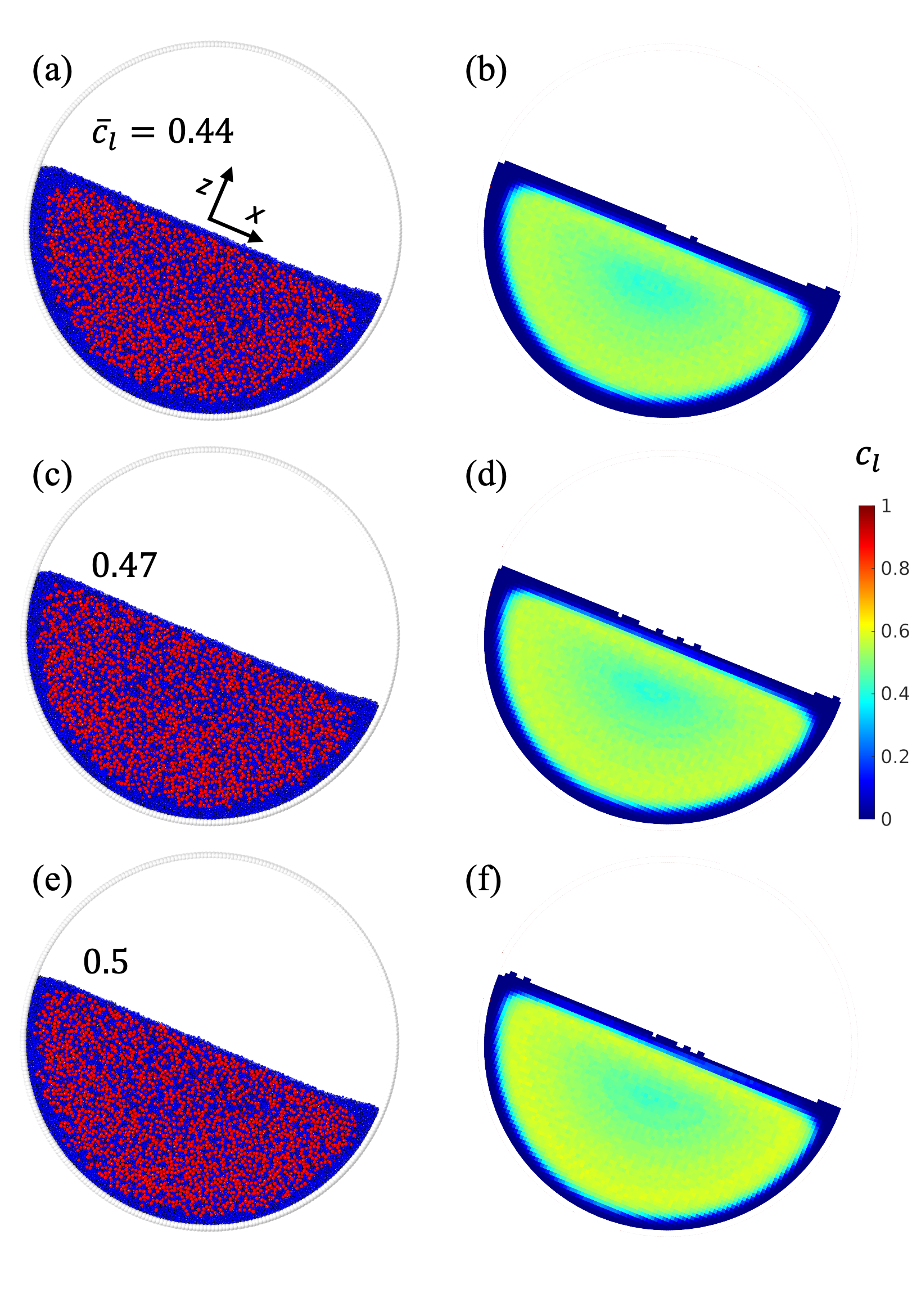}
\caption{
Segregation at steady state for $R_\mathrm{d}=1.5$, $R_\mathrm{\rho}=4$, $\bar c_\mathrm{l}$ of (a) 0.44, (b) 0.47, and (c) 0.5 using finer particles with $r_0/d_\mathrm{l}=50$.
Other parameters are the same as in Figure~\ref{fig7}b.
(b,d,f) Time-averaged large particle concentration, $c_\mathrm{l}$.
}
\label{fig12}
\end{figure}

\begin{figure}[t]
\centering
\includegraphics[width=\columnwidth]{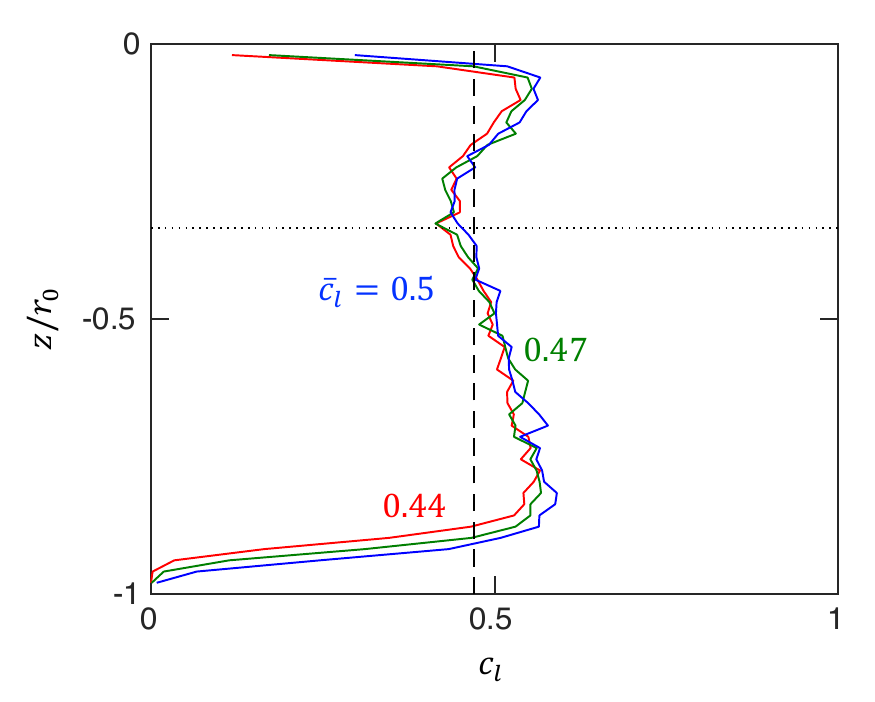}
\caption{Large-particle concentration $c_\mathrm{l}$ along a radial slice normal to the free surface in steady state for smaller particles with $r_0/d_\mathrm{l}=50$ at three different values of $\bar c_\mathrm{l}$.   
At the core center ($z/r_0=-0.3$, dotted vertical line), $c_\mathrm{l}\approx c_\mathrm{l,eq}$ (dashed horizontal line).  
 }
\label{fig13}
\end{figure}

Two significant results are evident  in Figures~\ref{fig9}-\ref{fig11}.
First,  segregation in a rotating tumbler  is significantly reduced for $\bar c_\mathrm{l}=c_\mathrm{l,eq}$, much like the earlier case of bounded heap flows; second, the particle concentration in the  core of the rotating tumbler bed saturates at  approximately the equilibrium concentration regardless of $\bar c_\mathrm{l}$.
Similar results are also observed  for the same size and density ratios ($R_\mathrm{d}=1.5$, $R_\mathrm{\rho}=4$) but with smaller absolute particle diameters (i.e., both $d_\mathrm{l}$ and $d_\mathrm{s}$ are scaled by a factor of two such that $r_0/d_\mathrm{l}=50$), shown in Figure~\ref{fig12}.
Particles remain mixed at steady state throughout most of the particle bed, indicating that the relative size of the tumbler compared to the particles is not crucial to the particles remaining mixed at $c_\mathrm{l}=c_\mathrm{l,eq}$.  
What is more interesting in this example is that the particles remain reasonably well-mixed not only for $\bar c_\mathrm{l}=c_\mathrm{l,eq}$ in Figure~\ref{fig12}c,d, but also for $\bar c_\mathrm{l}=c_\mathrm{l,eq}- 0.03$ in Figure~\ref{fig12}a,b and $\bar c_\mathrm{l}=c_\mathrm{l,eq}+ 0.03$ in Figure~\ref{fig12}e,f.
Radial concentration profiles for these three cases are shown in Figure~\ref{fig13}.
The nearly identical concentration profiles indicates that small deviations of $\bar c_\mathrm{l}$ from $c_\mathrm{l,eq}$  have little influence on the overall mixing of the particles in rotating tumblers.
Moreover, the concentration profile for $\bar c_\mathrm{l}=0.47$ in Figure~\ref{fig11} for $r_0/d_\mathrm{l}=25$ is nearly identical to those in Figure~\ref{fig13} for $r_0/d_\mathrm{l}=50$.  
This is because the particle concentration fields at steady state result from the balance between diffusion flux and segregation flux in the flowing layer. 
Since both fluxes are proportional to particle diameter,\cite{fan2014modelling} the effects of particle diameter cancel, and the concentration profiles are independent of $r_0/d_\mathrm{l}$, resulting in the green curve in Figure~\ref{fig13} matching the green curve in Figure~\ref{fig11} except near the surface and the corresponding region on the tumbler wall.
One difference between $r_0/d_\mathrm{l}=25$ and $r_0/d_\mathrm{l}=50$ is that the layers of nearly pure small particles (small $c_\mathrm{l}$) at the flowing layer surface and near the tumbler wall are thinner for $r_0/d_\mathrm{l}=50$ suggesting that the phenomenon driving this is related to the size of the particles compared to that of the tumbler.

\begin{figure}[t]
\centering
\includegraphics[width=\columnwidth]{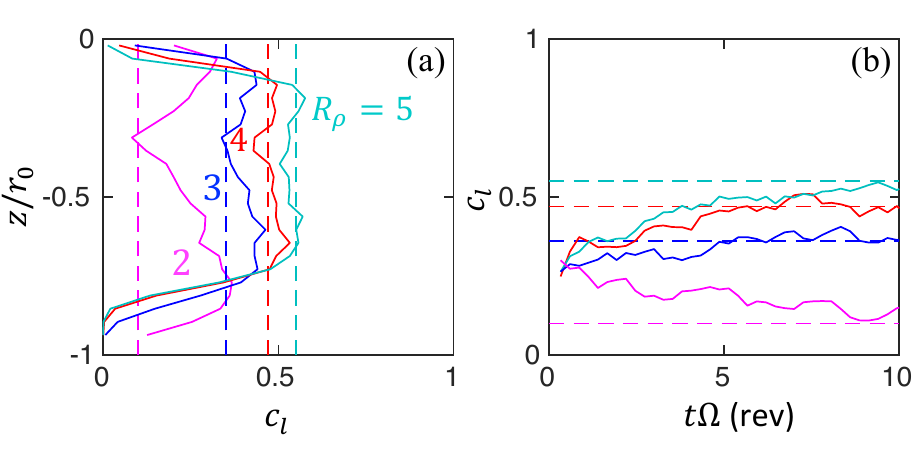}
\caption{For differing $R_\mathrm{\rho}$ and fixed global large-particle concentration $\bar{c}_\mathrm{l}=0.3$, the local large-particle concentration, $c_\mathrm{l},$ at the center of the tumbler core (i.e., $-0.32<z/r_0<-0.28$) is similar to the corresponding local equilibrium concentration $c_\mathrm{l,eq}=0.10$, 0.35, 0.47, and 0.55 \textcolor{blue} (dashed lines) for $R_\mathrm{d}=1.5$. (a) $c_\mathrm{l}$ along a radial slice normal to the free surface in steady state.
(b) $c_\mathrm{l}$ at the core center ($z/r_0=-0.3$) vs.\ number of rotations. }
\label{fig14}
\end{figure}

\begin{figure}[t]
\centering
\includegraphics[width=\columnwidth]{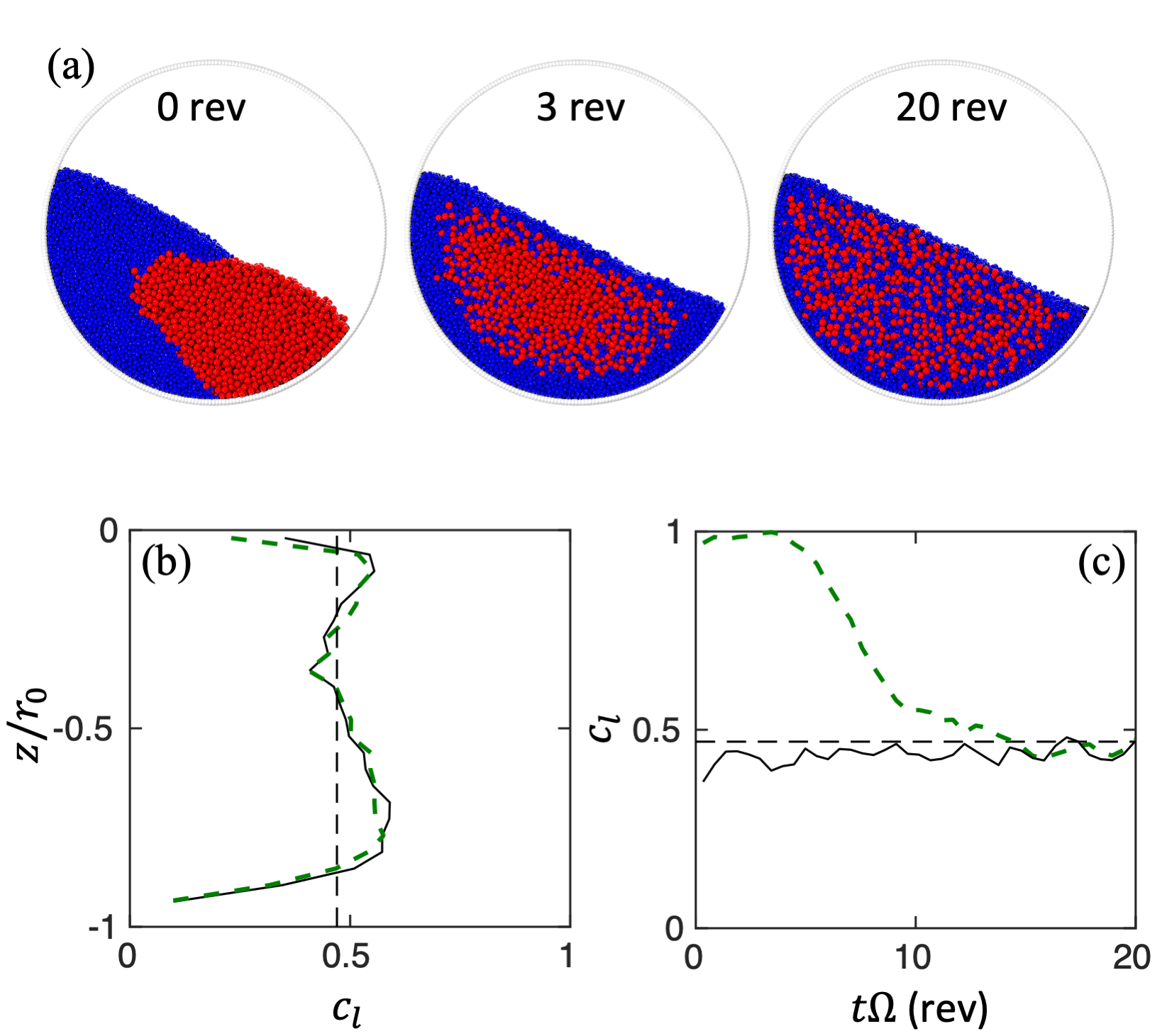}
\caption{(a) Initially fully segregated large heavy ($d_\mathrm{l}=3\,$mm, $\rho_\mathrm{l}=4\,$g/cm$^3$, red) and small light ($d_\mathrm{s}=2\,$mm, $\rho_\mathrm{s}=1\,$g/cm$^3$, blue) particles ($R_\mathrm{d}=1.5$, $R_\mathrm{\rho}=4$) with $\bar c_\mathrm{l}=c_\mathrm{l,eq}=0.47$ become well mixed at steady state. 
Left to right: initial segregated particle distribution, transient state after 3 rotations,  and steady state at 20 rotations. 
(b) Average steady state ($t \Omega>15$) radial large-particle concentration profile normal to free surface for well-mixed (solid curve) and fully segregated (dashed curve) initial conditions. Vertical dashed line corresponds to $c_\mathrm{l,eq}=0.47$.
(c) $c_\mathrm{l}$ at the core center ($z/r_0=-0.3$) vs.\ number of rotations. Horizontal dashed line corresponds to $c_\mathrm{l,eq}=0.47$.
}
\label{fig15}
\end{figure}

Returning now to the result in Figures~\ref{fig10} and \ref{fig12} showing that $c_\mathrm{l}\approx c_\mathrm{l,eq}$ in the  core of the tumbler bed regardless of the global concentration of large particles, we consider a different scenario in which the global concentration of large particles is held constant and $R_\mathrm{\rho}$ is varied. 
Figure~\ref{fig14} compares the radial concentration profiles and time evolution of $c_\mathrm{l}$ at the core center with $R_\mathrm{d}=1.5$ and  $\bar c_\mathrm{l} =0.3$, for different values of $R_\mathrm{\rho}$. 
In steady state $c_\mathrm{l}$ is close to $c_\mathrm{l,eq}$ (dashed lines) at the core center ($z/r_0\approx-0.3$) for the corresponding ($R_\mathrm{d}$, $R_\mathrm{\rho}$) for all four values of $R_\mathrm{\rho}$, as shown in Figure~\ref{fig14}a.  
This occurs regardless of whether $c_\mathrm{l}$ is increasing relative to $\bar c_\mathrm{l}$ to reach $c_\mathrm{l,eq}$, as is the case for $R_\mathrm{\rho}=2$ in Figure~\ref{fig14}b, or decreasing from $\bar c_\mathrm{l}$ to $c_\mathrm{l,eq}$, as is the case for $R_\mathrm{\rho}=3$, 4, 5. However, for cases with $\bar c_\mathrm{l}$ further away from $c_\mathrm{l,eq}$ as determined by the value of $R_\mathrm{\rho}$, it takes a longer time to reach equilibrium.

The appearance of  the equilibrium concentration in the core independent of $\bar c_\mathrm{l}$ appears to be a characteristic of steady-state segregation in tumbler flows.  
Apparently, as the particle mixture adjusts to its steady-state distribution, the core concentration relaxes to $c_\mathrm{l,eq}$ while the ``excess'' particles (small particles for $c_\mathrm{l}<c_\mathrm{l,eq}$ and large particles for $c_\mathrm{l}>c_\mathrm{l,eq}$)  are displaced to tumbler periphery where their concentration is enhanced.  
An explanation is as follows. Starting from a fully mixed condition, during the initial transient upward segregating particles rise in the flowing layer until the concentration in the lower portion of the flowing layer achieves the equilibrium concentration, $c_\mathrm{l,eq}$.  Once $c_\mathrm{l,eq}$ is established locally at the bottom of the flowing layer, no further segregation occurs in this region.  Since this layer with concentration $c_\mathrm{l,eq}$ is at the bottom of the flowing layer, it is deposited on the fixed bed at the core, while the particles that have segregated upward in the flowing layer are carried to the periphery of the tumbler to be deposited there.
This process is reinforced over several rotations until steady-state is reached, which corresponds to a concentration near $c_\mathrm{l,eq}$ in the core.  
We return to this phenomenon in the next section as a potential method to experimentally determine the equilibrium concentration for arbitrary binary particle mixtures.

One further consideration is that for the simulations described to this point, particles are initially well-mixed. 
However, in many practical industrial situations the two particle species are often initially segregated.
Figure~\ref{fig15} shows the mixing of the same particle mixture in Figure~\ref{fig9}b but starting from an initially fully segregated state  shown in Figure~\ref{fig15}a at the onset of rotation.  After three rotations (18\,s) the system forms a core of large particles with small particles at the periphery.  
After twenty rotations (120\,s), despite the initial complete segregation, particles are well mixed at steady state, just as occurs for the initially mixed case. 
Figure~\ref{fig15}b shows that the steady state concentration profiles are identical for the completely segregated and well-mixed initial  conditions.
However, the initially segregated case takes longer to reach equilibrium (15 rotations, $t\Omega \approx 15$, or 90\,s), as shown in Figure~\ref{fig15}c.
For the well-mixed case with $\bar c_\mathrm{l}= c_\mathrm{l,eq}=0.47 $, the mixture almost immediately reaches $c_\mathrm{l,eq}$, although for $\bar c_\mathrm{l} \neq c_\mathrm{l,eq}$ it takes about 5 rotations ($t\Omega \approx 5$, or 30\,s), as shown in Figure~\ref{fig11}b.
The results in Figure~\ref{fig15} suggest that initial segregation conditions of the mixture or variations in the initial local concentration, with sufficient time,  have little influence on the steady state segregation pattern, which always exhibits a concentration in the core that is close to $c_\mathrm{l,eq}$.

\section{Method to determine equilibrium concentration}

\begin{figure*}[t]
\centering
\includegraphics[width=\textwidth]{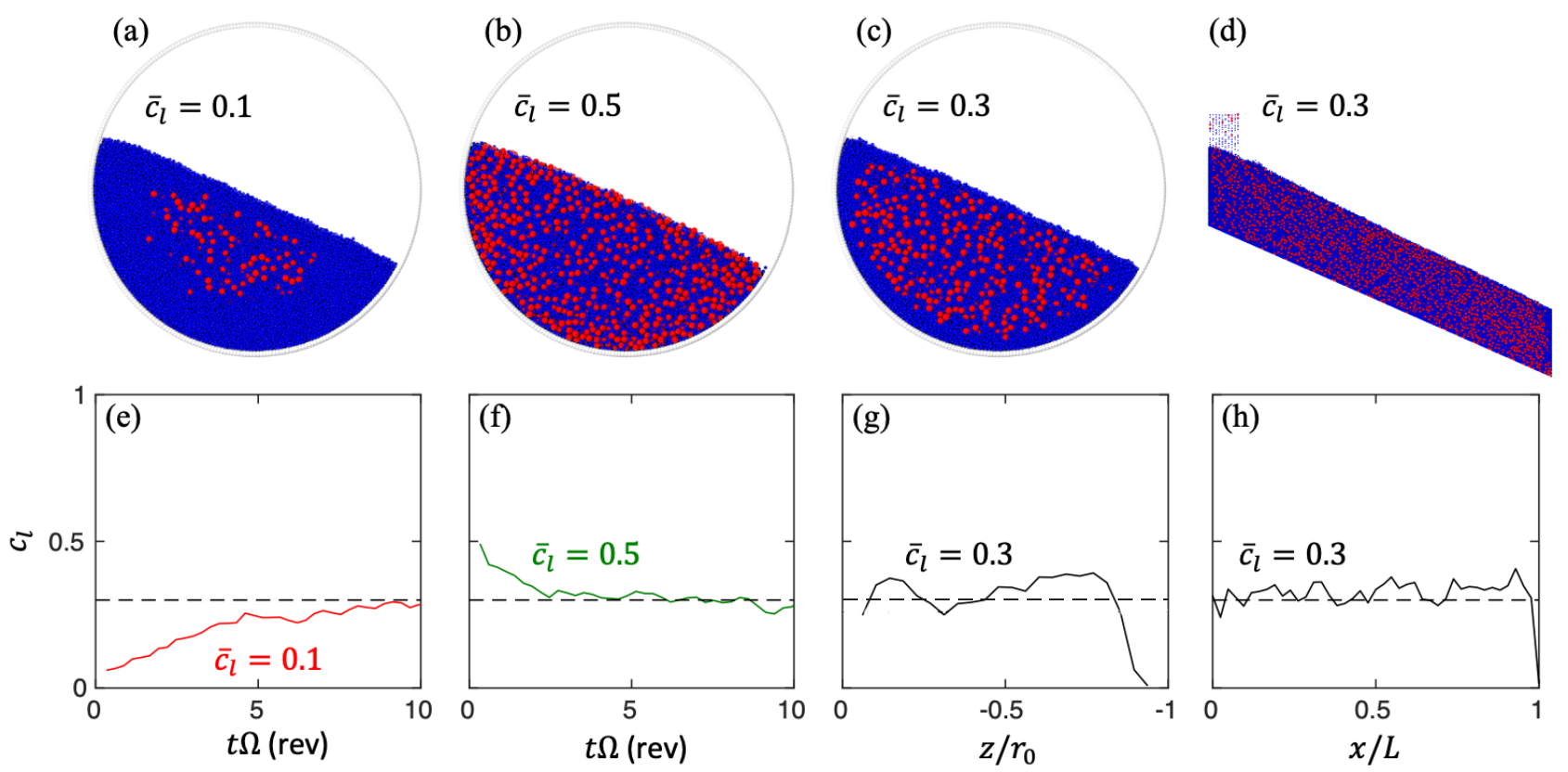}
    \caption{Simulations demonstrating practical determination of $c_\mathrm{l,eq}$ in a rotating tumbler for different initial mixture concentrations of (a,e) $\bar c_\mathrm{l}=0.1$ and (b,f) $\bar c_\mathrm{l}=0.5$ by measuring the core concentration value in steady state. The measured value of $c_\mathrm{l,eq}=0.3$ [dashed line in (e-h)] is tested in (c,g) a tumbler flow and (d,h) a bounded heap flow, both of which show minimal segregation. Note that the concentration profile in (g) is rotated $90^\circ$ from its orientation in previous figures so it can be easily compared to (h).}
\label{fig16}
\end{figure*}

The appearance of $c_\mathrm{l,eq}$ in the core of the rotating tumbler bed suggests a simple methodology to experimentally determine $c_\mathrm{l,eq}$ for an arbitrary binary particle mixture. 
The procedure is as follows.  Fill a rotating tumbler to half full with an arbitrary concentration binary mixture and tumble until steady state segregation is reached.  Remove a sample volume of particles from the core.
The measured $c_\mathrm{l}$ for the sample should be close to $c_\mathrm{l,eq}$. 
Then prepare a new mixture at the measured core concentration.  
It may be necessary to repeat these steps one or two times until no further variation in $c_\mathrm{l}$ at the core is observed.

The advantage of this approach is that the sizes and densities of the two particle species need not be known in advance, and, more significantly, no analog to Figure~\ref{fig3} is necessary (which requires many experiments or simulations). This makes the approach applicable for particles that may vary in properties other than just size or density, such as shape.~\cite{jones2021predicting} Using this approach, a few simple experiments can determine the equilibrium concentration that minimizes segregation.

Figure~\ref{fig16} shows DEM simulations demonstrating this approach for determining the equilibrium concentration for example mixtures with $R_\mathrm{d}= 2$ and $R_\mathrm{\rho}= 3$, starting with either of two different initial global concentrations: $\bar c_\mathrm{l}=0.1$ (Figure~\ref{fig16}a) or 0.5 (Figure~\ref{fig16}b). 
Figures~\ref{fig16}e,f show that $c_\mathrm{l}$ at the core center saturates at about 0.3 (dashed line) for both cases, indicating that $c_\mathrm{l,eq}\approx 0.3$, which is close to $c_\mathrm{l,eq}= 0.26$ for $R_\mathrm{d}= 2$ and $R_\mathrm{\rho}= 3$ determined from Figure~\ref{fig3}. 
Noting that small deviations in the concentration from the actual value of $\bar c_\mathrm{l,eq}$ are inconsequential (see Figure~\ref{fig12}), mixtures prepared at the measured $\bar c_\mathrm{l}= 0.3$ are tested in both the rotating tumbler and the bounded heap. Figures~\ref{fig16}c,d,g,h show that particles are relatively well-mixed over the two domains. Thus, for these parameters, only one iteration of the rotating tumbler experiment is necessary to narrow in on the approximate value of $c_\mathrm{l,eq}$, whether starting from too low of a concentration (Figures~\ref{fig16}a,e) or too high of a concentration (Figures~\ref{fig16}b,f).
The second iteration, Figures~\ref{fig16}c,g confirms the value for $c_\mathrm{l,eq}$.
These results indicate that this simple practical approach for finding the equilibrium concentration for arbitrary bidisperse particle mixtures holds substantial promise.

\begin{table}[bt]
  \begin{center}
  \begin{tabular}{lccc}
  \\
  \hline
  \headrow
      material  & color   &   diameter\,(mm) & density\,(g/cm$^3$)  \vspace{1mm}   \\
      \hiderowcolors
      steel  & dark & 3.14 $\pm$ 0.01 & 7.85 $\pm$ 0.02\\
      glass   & light & 1.54 $\pm$ 0.08 & 2.50 $\pm$ 0.05\\
        \hline
  \end{tabular}
  \caption{Particle properties in experiments: diameter from caliper measurements of 300 particles; density from the mass of 3 sets of 100 particles and average particle diameter.}
  \label{table1}
  \end{center}
\end{table}

\section{experiment}

\begin{figure}[t]
\centering
\includegraphics[width=0.95\columnwidth]{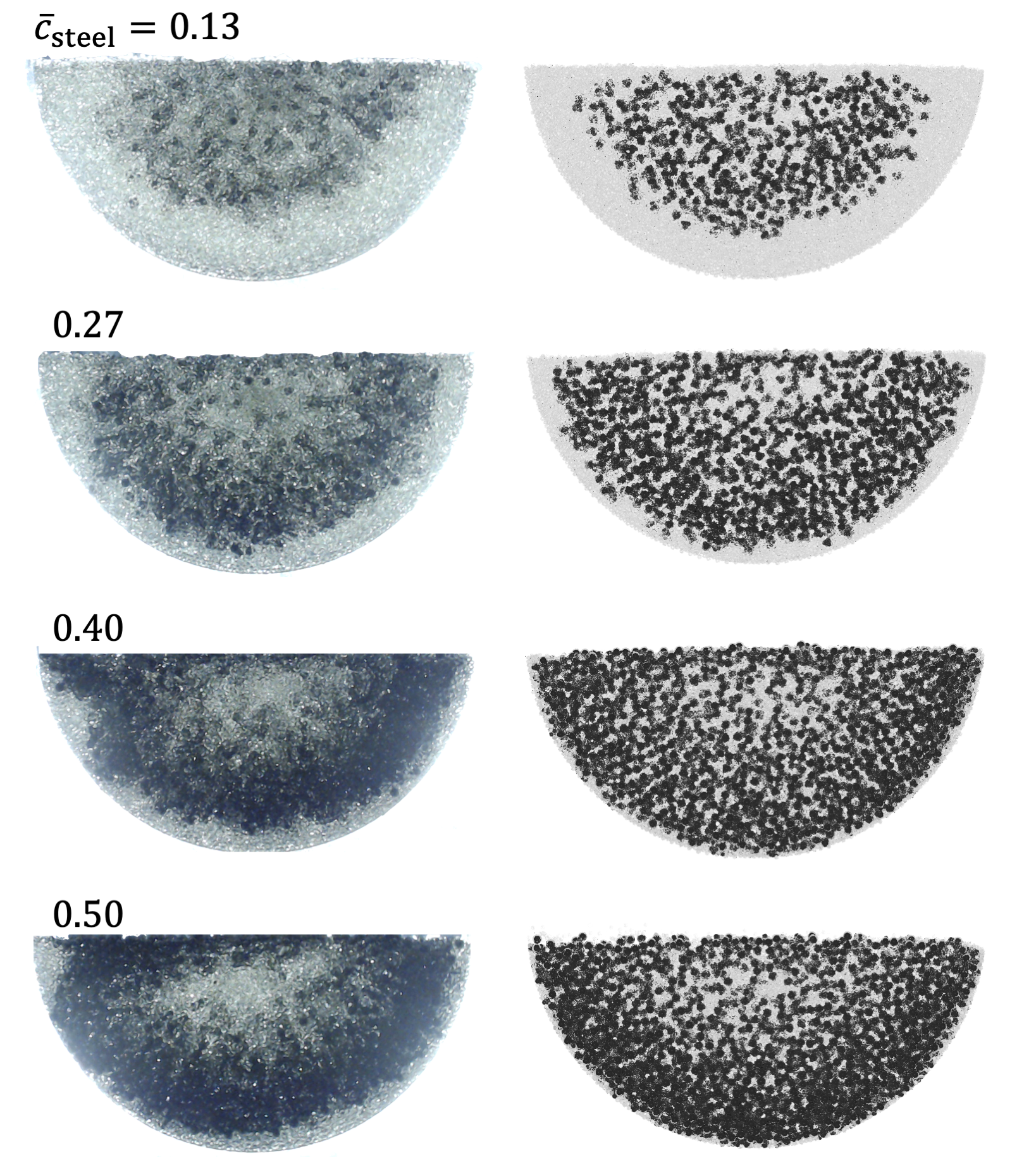}
\caption{Segregation of large steel (dark) and small glass (bright) particles in a 15\,cm rotating tumbler from experiments (left) and DEM simulations (right) with particle properties specified in Table~\ref{table1}.
$R_\mathrm{d}=2.04$ and $R_\mathrm{\rho}= 3.14$ for which $c_\mathrm{steel,eq} \approx 0.27$ according to Figure~\ref{fig3}.
Images from experiment and visualization of the simulations are obtained by backlighting one of the tumbler endwalls.
}
\label{fig17}
\end{figure}

Results to this point are based on DEM simulation, which for dense flows is usually quite accurate and relatively insensitive to particle and boundary interaction parameters.
To validate the simulations, experiments are performed in a rotating tumbler ($\Omega=4.08$~rev/min) with large steel and small clear glass spherical particles. The experimental set-up is similar to that in the simulation in Figure~\ref{fig9}, except that the 15\,cm diameter tumbler has two flat frictional endwalls made of clear acrylic and separated by 1.2\,cm ($3.82d_\mathrm{l}$) rather than the periodic endwall condition used in the simulations. One endwall is backlit to aid visualization of the particle distribution through the bulk. With the particle properties given in Table~\ref{table1}, the size ratio of large steel to small glass particles is $R_\mathrm{d}= 2.04$ and the density ratio $R_\mathrm{\rho}= 3.14$. The equilibrium concentration according to Figure~\ref{fig3} is $c_\mathrm{steel,eq}\approx 0.27$.

Figure~\ref{fig17} shows images of the segregation pattern from experiment for different global concentrations of large steel particles $\bar c_\mathrm{steel}$ after 10 minutes of rotation, when the segregation pattern has reached a steady state, as well as DEM results for the same conditions as the experiment.  
In this case, the DEM simulations include frictional endwalls with $\mu=0.5$ separated by 1.2\,cm, rather than using a periodic boundary condition as in the other tumbler simulations in this paper.

Although the visualization of the experimental results are less clear than the DEM simulation results, it is evident that the fundamental character of the particle distribution changes from mostly small glass particles surrounding a relatively mixed core at $\bar c_\mathrm{steel}=0.13$ to a generally mixed condition throughout the tumbler at $\bar c_\mathrm{steel}=0.27$ to mostly large steel particles surrounding a somewhat mixed core at $\bar c_\mathrm{steel}=0.40$ and $\bar c_\mathrm{steel}=0.50$.  
Focusing first on $\bar c_\mathrm{steel}=0.27$, which is at the predicted equilibrium concentration from Figure~\ref{fig3}, a band of small particles appears at the tumbler periphery for both experiments and simulations, confirming that this phenomenon occurs not only in the simulations but also in the experiments. 
That said, the degree of mixing is substantially better for $\bar c_\mathrm{steel}=0.27$ than for the lower concentration ($\bar c_\mathrm{steel}=0.13$), where the band of small particles at the periphery is wide and quite pure, or higher concentrations  ($\bar c_\mathrm{steel}=0.40$, 0.50), where the mixture is dominated by large particles in a wide band near the periphery, although a narrow band of small particles immediately adjacent to the tumbler wall persists in the experiments.  
Thus, while more work is needed to resolve these discrepancies, the general result of a more mixed state appearing for the equilibrium concentration than other concentrations appears to be the case for these tumbler experiments.

\section{discussion}
\label{discussion}

From both the DEM simulation results for rotating tumblers with $\bar c_\mathrm{l}\approx c_\mathrm{l,eq}$ (Figs.~\ref{fig9}-\ref{fig16} and the images from experiment ($\bar c_\mathrm{steel}=0.27$) in Figure~\ref{fig17}b, it is evident that the large particle concentration decreases sharply near the tumbler wall such that the overall mixing is imperfect even at the equilibrium concentration (i.e., $\bar c_\mathrm{l}=c_\mathrm{l,eq}$).
As mentioned earlier, the reduced large particle concentration near the tumbler wall has been noted previously \citep{pereira2014segregation,thomas2018evidence} but not explained.  
A possible explanation is that the equilibrium concentration in Figure~\ref{fig3} derived from depth-averaging heap flow segregation flux data as shown in Figure~\ref{fig2} may not be universal and could additionally vary locally with depth in the flowing layer.
Specifically, our recent study on the forces on a single intruder particle in shear flow demonstrates that the equilibrium condition for single intruders depends on the local shear rate gradient and pressure.\cite{jing2021unified}
Given that in free surface flows the streamwise particle velocity  exponentially decreases with depth \cite{fan2013kinematics} and the local lithostatic pressure linearly increases with depth for a uniform density profile, it is reasonable to infer that the equilibrium concentration associated with non-mixing conditions may vary locally with depth in the flowing layer. As a result, local segregation may occur despite the global concentration satisfying the equilibrium condition. 
Such local segregation appears to be negligible in heap flows, as the segregation is not fully developed before the particles deposit onto the fixed bed. However, in rotating tumblers, with particles repeatedly segregating in the flowing layer until reaching equilibrium, the variations in the local shear rate gradient and lithostatic pressure through the depth of the flowing layer may have sufficient time to result in greater segregation than in heap flows. This is evident in the concentration profile for $\bar c_\mathrm{l}=c_\mathrm{l,eq}$ in Figure~\ref{fig16}g for a rotating tumbler deviating more from $c_\mathrm{l,eq}$ than that for a bounded heap flow in Figure~\ref{fig16}h.

A second reason for reduced large-particle concentration near the rotating tumbler wall may be the complex flow kinematics at the end of the downstream portion of the flowing layer where particles interact with the cylindrical tumbler wall before entering solid body rotation. 
The thickness of the small particle band near the tumbler wall decreases somewhat for simulations using smoother tumbler walls, suggesting that particle-wall interactions are important.
The effect of the tumbler wall roughness is consistent with previous results indicating how the wall roughness can affect secondary flows in cylindrical tumblers \citep{d2022mechanisms} and both secondary flows and segregation in spherical tumblers \citep{d2015influence,d2016influence} even far from the tumbler walls.

Despite the visible small particle band at the tumbler wall, evident visually in Figure~\ref{fig9}b and in the concentration profile in Figure~\ref{fig11}, it is clear that segregation is largely suppressed for $\bar c_\mathrm{l}= c_\mathrm{l,eq}=0.47$ in this case.
In contrast, for $\bar c_\mathrm{l}=0.3$, $c_\mathrm{l}$ decreases nearly to zero at the surface of the flowing layer and to zero (pure small particles) in a band at the tumbler periphery.  
For $\bar c_\mathrm{l}=0.7$, $c_\mathrm{l}$ increases at the surface of the flowing layer and at the periphery.
Thus, despite the deviation of $c_\mathrm{l}$ from a uniform value for $\bar c_\mathrm{l}= c_\mathrm{l,eq}$ at the top of the flowing layer and near the cylindrical tumbler wall in this specific case as well as those in (Figs.~\ref{fig12}-\ref{fig16}), most particles in the tumbler remain relatively well-mixed at the equilibrium concentration, whereas at other concentrations the particles tend to be substantially more segregated.

\section{Conclusion}

For particle mixtures varying simultaneously in size and density, the two corresponding segregation mechanisms (percolation and buoyancy, respectively) interact with each other resulting in segregation behavior significantly different from size or density segregation alone. In particular, mixtures of large heavy and small light particles can exhibit an equilibrium concentration, $c_\mathrm{l,eq}$, at which the two segregation mechanisms are locally balanced and the net segregation flux is zero. This leads to a methodology in which a particle system can be designed to reduce or even prevent segregation by specifying the optimal combination of particle size ratio $R_\mathrm{d}$, density ratio $R_\mathrm{\rho}$, and overall mixture concentration $\bar c_\mathrm{l}$.

The near overlap of the equilibrium concentration curves in Figure~\ref{fig7} with the non-segregating region determined from heap flow simulations demonstrates not only  the accuracy of the equilibrium conditions predicted in Figure~\ref{fig3} but also the potential for designing minimally-segregating granular mixtures by feeding particles initially mixed at the equilibrium concentration, i.e. $\bar c_\mathrm{l}=c_\mathrm{l,eq}$.
This potential to intentionally design minimally-segregating mixtures also extends to bidisperse particle mixtures in rotating tumblers with global concentration $\bar c_\mathrm{l}=c_\mathrm{l,eq}$, even when particles are initially segregated.  
In addition, particles in the tumbler core tend toward the equilibrium concentration even when $\bar c_\mathrm{l}\neq c_\mathrm{l,eq}$.
This leads to a methodology to experimentally find $ c_\mathrm{l,eq}$ using a rotating tumbler by measuring the concentration in the core.

Given the similarities in flow kinematics among heap, rotating tumbler, and other geometries where a thin surface layer of particles flows relative to the bulk, we expect that the equilibrium condition in Figure~\ref{fig3} is generally applicable to surface flows in other geometries.\cite{deng2020modeling,isnergranular,isner2020axisymmetric}
However, the equilibrium condition in Figure~\ref{fig3} may not be universal and may change somewhat with flow kinematics, since the solid volume fraction, overburden pressure, shear rate, and shear rate gradient affect particle segregation.
For example, the equilibrium condition for equal-volume particle mixtures under vibration is $R_\mathrm{d}\approx R_\mathrm{\rho}$,\cite{hong2001reverse} which clearly differs from the $c_\mathrm{l,eq}=0.5$ curve in Figure~\ref{fig3}.
Likewise, we expect that the equilibrium concentration would be different in wall-driven shear flow between two planes (due to the overburden pressure) \cite{fry2018effect} or in chute flow where substantial slip can occur at the base of the flow or the velocity exhibits a Bagnold-type profile.
Nevertheless, our results demonstrate the potential for specifying particular combinations of particle size ratio, density ratio, and concentration to promote mixing and minimize segregation in granular surface flows.
Future studies connecting equilibrium conditions to flow kinematics  and achieving a phase diagram analogous to Figure~\ref{fig3} for other flows such as chute flows and boundary-driven (wall or intruder) flows are  warranted.

\section*{acknowledgements}
We gratefully acknowledge helpful discussions with Yi Fan, John Hecht, and Lu Jing.
This material is based upon work supported by the National Science Foundation under Grant No.~CBET-1929265.

%
%
%


\bibliography{main}



\end{document}